\documentclass[apj,twocolumn]{openjournal}
\usepackage{amsmath}
\usepackage{booktabs}
\usepackage{multirow}
\usepackage{color}
\usepackage{soul}
\usepackage{orcidlink}

\usepackage[dvipsnames]{xcolor} 

\newcommand{\spectral}[3]{#1\,{\sc #2}\,{$\lambda #3$}}

\usepackage[breaklinks,colorlinks,citecolor=blue,urlcolor=blue]{hyperref}

\defcitealias{Green2018}{G18}
\defcitealias{Rodriguez2023}{R23}
\defcitealias{vanRoestel2022}{vR22}

\renewcommand{\arraystretch}{1.2}  
\usepackage{listings}
\usepackage{color}
\definecolor{dkgreen}{rgb}{0,0.6,0}
\definecolor{gray}{rgb}{0.5,0.5,0.5}
\definecolor{mauve}{rgb}{0.58,0,0.82}
\definecolor{golden}{rgb}{0.86,0.65,0.01}
\begin{document}


\title{JWST observations of three long-period AM CVn binaries: detection of the donors and hints of magnetically truncated disks}

\author{\vspace{-30pt}Kareem El-Badry\,\orcidlink{0000-0002-6871-1752}$^{1}$}
\author{Antonio C. Rodriguez\,\orcidlink{0000-0003-4189-9668}$^{2,1}$}
\author{Matthew J. Green\,\orcidlink{0000-0002-0948-4801}$^{3,4}$}
\author{Kevin B. Burdge\,\orcidlink{0000-0002-7226-836X}$^{5}$}

\affiliation{$^1$Department of Astronomy, California Institute of Technology, 1200 E. California Blvd., Pasadena, CA 91125, USA}
\affiliation{$^2$Center for Astrophysics $|$ Harvard \& Smithsonian, 60 Garden Street, Cambridge, MA 02138, USA}
\affiliation{$^3$Homer L. Dodge Department of Physics and Astronomy, University of Oklahoma, 440 W. Brooks Street, Norman, OK 73019, USA}
\affiliation{$^4$JILA, University of Colorado and National Institute of Standards and Technology, 440 UCB, Boulder, CO 80309-0440, USA}
\affiliation{$^5$Department of Physics, Massachusetts Institute of Technology, Cambridge, MA 02139, USA}

\email{Corresponding author: kelbadry@caltech.edu}

\begin{abstract}
We present {\it JWST}/NIRSpec  high-cadence infrared  spectroscopy of three long-period, eclipsing AM CVn binaries, Gaia14aae, SRGeJ0453, and ZTFJ1637. These systems have orbital periods of 50–62 minutes and cool donors that are undetectable in the optical. The data cover a wavelength range of 1.6–5.2 $\mu$m at resolution $R=1000-2000$. We obtained 150-200 spectra of each system over two orbits, split between the G235M and G395M gratings. All three systems show strong, double-peaked He I emission lines dominated by an accretion disk. These lines are nearly stationary but contain radial velocity (RV) variable sub-components that trace  stream-disk interactions. In Gaia14aae and SRGeJ0453, we detect two Na I doublets in emission whose RVs track the irradiated face of the donor, marking the first direct detection of the donors of long-period AM CVns. No absorption lines from the donors are detected, implying that the IR excesses observed in many long-period AM CVns primarily trace disks, not donors. The He I emission profiles in all systems lack high-velocity wings and show no emission beyond $\approx 1500 \rm  km\,s^{-1}$. The morphology of the disk eclipses and Doppler tomograms are best reproduced by models in which the disk is truncated well outside the white dwarf and only material at $r \gtrsim 0.07\,R_{\odot}$ contributes to the disk emission. We interpret this as possible evidence of magnetized white dwarf accretors. For plausible mass transfer rates, the truncation radii imply surface magnetic fields of $B = 30-100$\,kG, consistent with recent constraints based on X-ray periodicity. The absence of cyclotron humps out to 5\,$\mu$m rules out stronger MG-level fields. We make the data from the program publicly available to the community.

\end{abstract}

\maketitle

\section{Introduction}
\label{sec:intro}
AM CVn binaries are ultra-short period  ($P_{\rm orb} \lesssim 70$ min) binaries containing a white dwarf (WD) accreting from a low-mass, helium-rich companion \citep[e.g.][]{Nelemans2005, Solheim2010, Green2025}. The companions (``donors'') are supported primarily by degeneracy pressure and have masses $M_2 \lesssim 0.1\,M_{\odot}$, making them helium-dominated analogs of brown dwarfs and giant planets. 

AM CVns are the terminal stage of stable mass transfer in a binary with a WD accretor. The initial state of the donor stars is uncertain, and several different models have been proposed: before mass transfer started, it may have been a helium WD \citep[e.g.][]{Paczynski1967}, a core helium burning star \citep[e.g.][]{Savonije1986}, or a subgiant with a helium core \citep[e.g.][]{Tutukov1985}. Candidate progenitor systems of all three models have been observed, but their relative importance is still uncertain. 

AM CVns can reach extremely short orbital periods: the shortest observed period is only 5 min \citep{Israel2002}, and for the double WD channel, models predict minimum periods of less than 1 min \citep[e.g.][]{Chen2022}. AM CVns with  $P_{\rm orb}\lesssim 20$ min have high mass transfer rates and hot accretion flows that make them bright in X-rays and in the UV \citep[e.g.][]{Ramsay2006}. They are also predicted to be among the loudest sources of millihertz gravitational waves in the Milky Way, and {\it LISA} is predicted to discover tens of thousands of them \citep[e.g.][]{Nelemans2004, Burdge2020, Kupfer2024}. 

As AM CVns evolve, the donors -- which were initially stars or WDs -- are whittled down by mass transfer. Their masses fall, first to brown dwarf masses, and eventually to a few Jupiter masses. Due to their degenerate equation of state, the donors expand as they are stripped down, causing the orbits to widen. Adiabatic expansion cools the donors to temperatures of order 1000\,K \citep[e.g.][]{Deloye2007}. The binaries' orbital evolution is dominated by gravitational wave radiation, perhaps with additional contributions due to magnetic braking. Gravitational waves become weaker as the donors lose mass and the orbits expand, reducing the mass transfer rate. As a result, long-period AM CVns ($P_{\rm orb} \gtrsim 45$ min) have faint disks that only very rarely undergo outbursts, with predicted recurrence timescales of years to centuries \citep[e.g.][]{Levitan2015}. The evolution of AM CVns slows as their orbits widen, and so a large majority of AM CVns are predicted to be found at long periods \citep[$P_{\rm orb} \gtrsim 60$\,min; e.g.,][]{Nelemans2001}. However, the infrequent outbursts and faint accretion disks of long-period AM CVns make them harder to detect observationally and underrepresented in observed samples \citep[e.g.][]{Green2025, Kara2025}.

Because they are so cold, the donor stars in long-period AM CVn binaries have never been observed. Their existence has been inferred only indirectly, from the presence of an accretion disk, and in some cases, from eclipses of the accreting WD \citep[e.g.][]{Green2018, vanRoestel2022}. With spectral energy distributions (SEDs) that are predicted to peak at 2-5 $\mu$m, AM CVn donors are best observed in the infrared (IR). The SEDs of most long-period AM CVns display IR excesses that have been interpreted as evidence of the donor \citep{Green2020, vanRoestel2022}. However, this interpretation has not been confirmed spectroscopically, and it is uncertain to what extent the accretion disk may also contribute in the IR. The only AM CVns with published IR spectroscopy are GP Com ($P_{\rm orb} = 46.6$ min) and V396 Hya ($P_{\rm orb} = 65.1$ min) \citep{Dhillon2000, Kupfer2016}. These systems are the nearest AM CVns known and are accessible at $1-2.4\,\mu$m with ground-based spectroscopy. Their near IR spectra revealed helium emission lines and no signatures of the donors. However, the donors are predicted to dominate the spectra only at longer wavelengths, which are inaccessible from the ground.

In this work, we present IR observations of three long-period AM CVns obtained with {\it JWST}. Our targets are all eclipsing systems that have been studied extensively in the optical and have robustly measured physical parameters from light curve modeling. Compared to ground-based observations, our data reach longer wavelengths, have higher sensitivity, and are unaffected by telluric absorption and sky emission, which are a significant challenge for ground-based IR spectroscopy. 

The remainder of this paper is organized as follows. In Section~\ref{sec:sample}, we summarize our targets, observations, and data reduction. Section~\ref{sec:res} contains our main results, including phase-averaged spectra (Section~\ref{sec:phase_avg}), phase variability (Section~\ref{sec:trailed}), eclipse mapping (Section~\ref{sec:eclipse_mapping}), constraints on magnetic fields (Section~\ref{sec:magnetic}), and constraints on thermal emission from the donors (Section~\ref{sec:thermal_donors}). We summarize our results in Section~\ref{sec:conclusions}.

\section{Targets and observations}
\label{sec:sample}

\begin{table*}[]
\begin{tabular}{lllllllllll}
Name & $P_{\rm orb}$ & $R_{\rm donor}$ & $M_{\rm WD}$ & $M_{\rm donor}$ & inc & $a$ & $T_{\rm eff,\,WD}$ & $G$ & $W_1$ & Reference \\
& [min] & [$R_{\odot}$] & [$M_{\odot}$] & [$M_{\odot}$] & [deg] & [$R_{\odot}$] & [K] & [mag] & [mag] & \\ 
\hline 
Gaia14aae     & 49.7  & $0.060\pm0.002$ & $0.87\pm0.02$         & $0.0250\pm0.0013$ & $86.3\pm0.3$   & $0.430\pm0.003$ & $12900\pm200$ & 18.29 & 17.29 & \citetalias{Green2018} \\
SRGeJ0453     & 55.1  & $0.078\pm0.012$ & $0.85^{+0.04}_{-0.05}$ & $0.044\pm0.020$   & $82.5\pm1.5$   & $0.460\pm0.008$ & $16570\pm250$ & 18.58 & 17.19 & \citetalias{Rodriguez2023} \\
ZTFJ1637      & 61.5  & $0.068\pm0.007$ & $0.90\pm0.05$         & $0.023\pm0.008$   & $82.7\pm0.09$  & $0.501\pm0.009$ & $11200\pm300$ & 19.35 & 18.69 & \citetalias{vanRoestel2022} \\        
\end{tabular}
\caption{Target list. $G$ magnitudes are from {\it Gaia} DR3 and $W_1$ magnitudes are from the unWISE catalog \citep{Schlafly2019}. Masses, radii, and inclinations were measured from optical light curve modeling, while WD effective temperatures were estimated from SED fitting. Parameter constraints are taken from the references listed in the last column. }
\label{tab:sample}
\end{table*}

\begin{table*}[]
\begin{tabular}{llllllllllll}
Name & RA & Dec & $\varpi$ & $P_{\rm orb}$ & $t_0$ & $F_{\rm irr}$ & $T_{\rm uni}$ & $T_{\rm day}$ & $T_{\rm night}$ \\
& [deg] & [deg] & [mas] & [d] & [BMJD TDB] & [$\rm erg\,s^{-1}\,cm^{-2}$] & [K] & [K] & [K] \\ 
\hline 
Gaia14aae     &  242.891522  &  63.142128 &  $3.90 \pm 0.12$         &   $0.0345195708$  & $57153.689097$  & $(7.30\pm0.80)\times10^{8}$ & $1339\pm27$ & $1439\pm29$ & $1266\pm25$ \\
SRGeJ0453     &  73.500612  &  62.412677 &  $4.28 \pm 0.17$ &   $0.0382501389$   & $59967.226794$     & $(1.83\pm0.27)\times10^{9}$ & $1683\pm54$ & $1808\pm58$ & $1591\pm51$ \\
ZTFJ1637      &  249.431259  &  49.294566 &  $4.89 \pm 0.20$         &  $0.042707771$    & $58370.23498$  & $(2.87\pm0.40)\times10^{8}$ & $1058\pm44$ & $1137\pm48$ & $1000\pm42$ \\        
\end{tabular}
\caption{Additional astrometric, orbital, and irradiation properties. $F_{\rm irr}$ and donor temperatures are predicted from WD irradiation (Section~\ref{sec:thermal_donors}; Equations~\ref{eq:irr}-\ref{eq:Tirr_eps}) assuming $A_B=0$ and $\varepsilon=0.8$.}
\label{tab:sample2}
\end{table*}

\subsection{Targets}

  We observed three eclipsing, long-period AM CVns whose properties are summarized in Tables~\ref{tab:sample} and~\ref{tab:sample2}. All three systems have been modeled extensively in the optical and have well-measured orbital ephemerides, mass ratios, and inclinations inferred from light curve modeling. The effective temperatures of the accreting WDs have been inferred in the literature by fitting the systems' SEDs; these estimates are more uncertain, because the relative flux contributions of the WD and accretion disk are uncertain. For Gaia14aae and ZTFJ1637, the radius of the primary WD was inferred from eclipse fitting, and this was translated to a mass constraint via a theoretical mass-radius relation. For SRGeJ0453, the primary WD radius and thus mass were constrained mainly via SED fitting, likely leading to larger systematic uncertainties. We briefly summarize the properties of the three systems below. 

\subsubsection{Gaia14aae}
Gaia14aae ({\it Gaia} DR3 source ID 1629388752470472704) is a fully-eclipsing AM CVn with an orbital period of 49.7 min. It was discovered following a series of outbursts in 2014 \citep{Campbell2015}, when it brightened from a quiescent magnitude of $V \approx 18.5$ to a maximum brightness of  $V \approx 13.5$. No comparable outbursts have been observed since then. The system has been studied with high-cadence optical photometry \citep[][hereafter G18]{Green2018} and spectroscopy \citep{Green2019}. The mass and radius constraints reported in Table~\ref{tab:sample} come from modeling of the high-cadence light curve by \citetalias{Green2018}, particularly the duration and phase of the bright spot eclipse. 

The measured radius of the donor is larger than predicted for a fully-degenerate object of the same mass, and \citetalias{Green2018} found that it is best matched by models for AM CVns formed from donors that initiated mass transfer as subgiants. An infrared excess is observed in the SED with a best-fit effective temperature of $2070\,{\rm K}$; this has been tentatively attributed to emission from the donor \citep{Macrie2024}. Finally, \citet{Maccarone2024} reported X-ray emission from the system that is modulated on the orbital period, which they interpreted as evidence that the accretion disk is truncated by the magnetic field of the accretor.  

\subsubsection{SRGeJ0453} 
SRGeJ0453 ({\it Gaia} DR3 source ID 477829370972112000) is a fully-eclipsing AM CVn with a period of 55.1 minutes. Discovered through a joint analysis of optical light curves from the Zwicky Transient Facility (ZTF) and X-ray fluxes from eROSITA \citep[][hereafter R23]{Rodriguez2023}, the system has never been observed to outburst. \citetalias{Rodriguez2023} noted a significant IR excess at wavelengths $\lambda \gtrsim 2\,\mu$m, which they attributed to the donor and/or disk. They inferred a WD effective temperature of $16570\pm 250\,{\rm K}$ by fitting the SED with a blackbody model. However, they found a lower effective temperature of $12200\pm 400\,{\rm K}$ and comparably good fit when fitting the SED with a DB WD model, so the reported uncertainties are likely underestimated. The source has a relatively hard X-ray spectrum and shows strong X-ray variability: it was undetected in two of the four eROSITA surveys while being well above the detection threshold in the other two. These X-ray properties led \citetalias{Rodriguez2023} to speculate that the WD may be magnetic. 

\subsubsection{ZTFJ1637} 
ZTFJ1637 ({\it Gaia} DR3 source ID 1410739870171621504) has an orbital period of 61.5 minutes and undergoes a grazing eclipse. The system was discovered via an optical light curve search for eclipses by \citep[][hereafter vR22]{vanRoestel2022} and has never been observed to outburst. \citetalias{vanRoestel2022} did not detect evidence of a bright spot in the system's optical spectra or photometry; they concluded that both the disk and bright spot contribute negligibly to the optical flux. The system also exhibits an IR excess, which \citetalias{vanRoestel2022} found to be well-explained by thermal emission from the donor. Like Gaia14aae, the donor has a larger inferred radius than predicted by double WD progenitor models; it can be better explained by helium star or evolved CV progenitors \citep[see also][]{Sarkar2023}.

\subsection{Observations}
\label{sec:obs}
Through JWST program 4979, we observed each system with NIRSpec for 1.05 orbital periods with each of the G235M and G395M gratings. We used the bright object time-series (BOTS) mode with the S1600A1 fixed slit and the SUB2048 subarray. We used the NRSRAPID readout pattern and 50 groups per integration, for an effective exposure time of 46 seconds. This setup yielded 75-88 spectra in each grating, with at least one full eclipse covered in each grating. The spectral resolution in both gratings increases toward redder wavelengths, ranging from $1000 \lesssim  R \lesssim 1800$ for G235M and $1000 \lesssim  R \lesssim 1500$ for G395M \citep{Shajib2025}. 

The G235M and G395M observations of each object were separated by time baselines ranging from one day to two months. The flux level in the overlap regions between the two gratings are consistent in all cases, suggesting that there was little change in the mass transfer rate between observations in the two gratings. 

\subsection{Data reduction}
\label{sec:data_reduction}

We reduced the data using \texttt{Eureka!}\ \citep{Bell2022}, which extends the JWST Science Calibration
Pipeline \citep[version 1.18.0;][]{Bushouse2025} with CRDS version \texttt{jwst\_1364.pmap}. The JWST pipeline performs bias and dark current subtraction, reference pixel correction, linearity corrections, flat-fielding, flagging of pixels affected by cosmic rays or bad pixels, and wavelength calibration. 

NIRSpec detectors have an evolving population of bad pixels, driven primarily by ``hot'' pixels with  higher than normal dark current. Not all such pixels are flagged as \texttt{DO\_NOT\_USE}: in many cases, a particularly hot pixel is flagged, but the four ``warm'' pixels bordering it are not. We inspected the 2D spectra and manually flagged about 100 groups of such warm pixels, as well as a few isolated pixels with high or strongly negative fluxes in all groups. Once these pixels are masked, the reduction pipeline interpolates their fluxes from spatially adjacent pixels in the 2D spectra. This removes the effects of most hot pixels, although a few that fall on the center of the trace cannot be fully corrected, since BOTS mode does not include dithering.  

We used \texttt{Eureka!}\ for background subtraction, order tracing, and optimal extraction of 1D spectra. The resulting 1D spectra are significantly more stable, as reflected in variations of the continuum flux from frame to frame, than the \texttt{x1dints} files produced by the automated reductions available on MAST. We subsequently performed absolute flux calibration using NIRSpec observations of the WD standard star G191-B2B obtained as part of JWST calibration program 7565, described by \citet{Gordon2022}. We reduced and extracted spectra of G191-B2B, which were obtained with the G235M and G395M gratings with the S1600A1 aperture, using \texttt{Eureka!}\ and the same settings used for our science targets. We then used the CalSpec \citep{Bohlin2007} model spectrum of G191-B2B to obtain a conversion between counts per second and flux in physical units as a function of position on the NRS1 detector, and finally multiplied the uncalibrated \texttt{Eureka!}\ spectra of the science targets by this position-dependent conversion factor. 

We estimate the per-exposure SNR empirically from the standard deviation of the continuum flux in regions without strong spectral lines. In the G235M grating, this yields a typical SNR of 4-5 per pixel for Gaia14aae and SRGeJ0453, and $\rm SNR\approx 3$ per pixel for ZTFJ1637. In the G395M grating, it yields a typical SNR of 3-4 per pixel for Gaia14aae and SRGeJ0453, and $\rm SNR\approx 2.5$ per pixel for ZTFJ1637. These empirical estimates are significantly lower than the values of $\rm SNR = 10-15$ per pixel implied by the formal uncertainties reported by both \texttt{Eureka!}\ and the JWST pipeline. Comparison of the three sources' phase-averaged spectra reveals correlated noise that is likely driven mainly by NIRSpec's large number of warm pixels, whose effects are amplified at low SNR and are not accounted for in the pipeline-reported uncertainties. 

Our final reduced data consist of 75-88 extracted 1D spectra for each system in each grating. These data, as well as the raw \texttt{uncal} and \texttt{rateints} files from the program, are available \href{https://doi.org/10.22002/jjsjy-37d32}{online}. 

\section{Results}
\label{sec:res}

\subsection{Phase-averaged spectra}
\label{sec:phase_avg}

\begin{figure}
    \centering
    \includegraphics[width=\columnwidth]{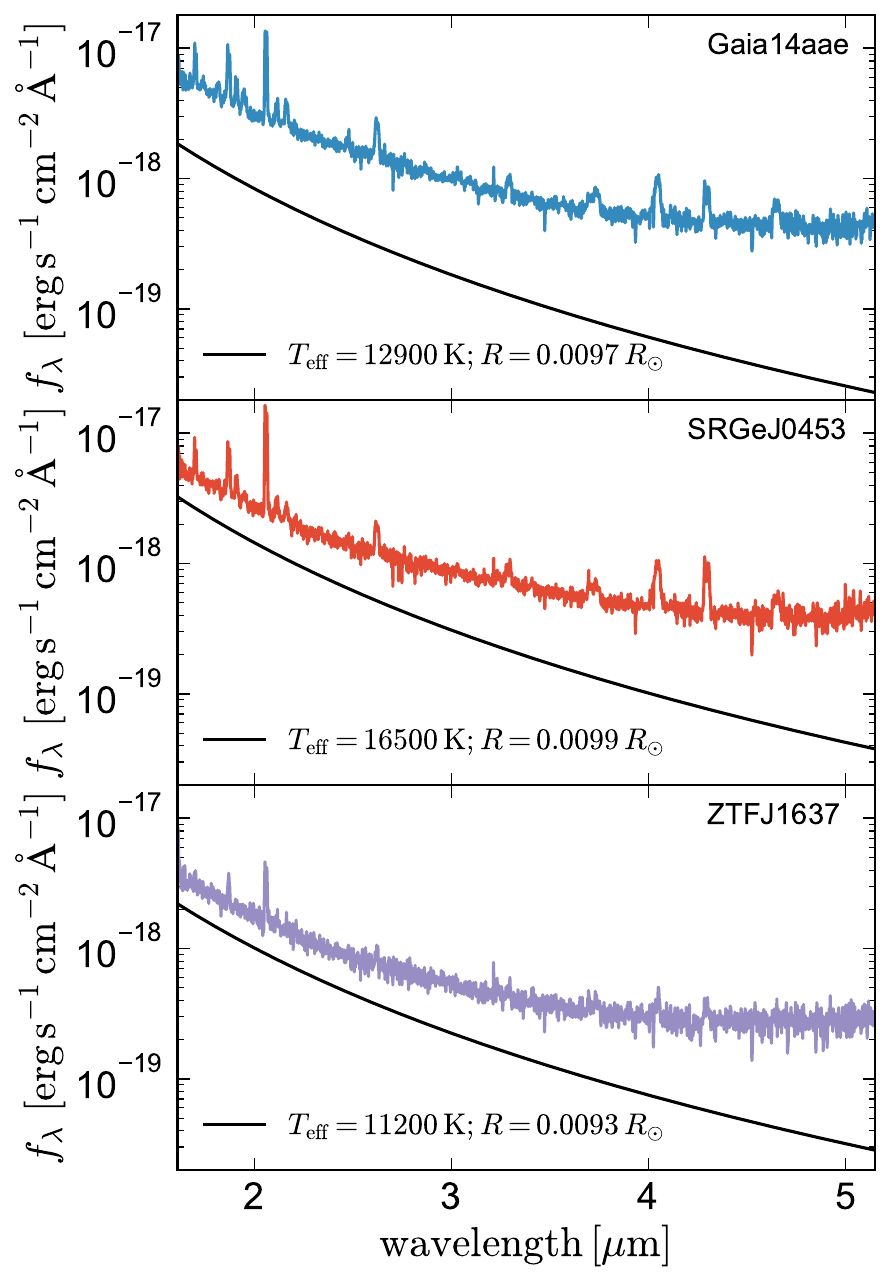}
    \caption{Phase-averaged spectra of all three sources, ordered by increasing orbital period. We combine data from the G235M and G395M observations and plot the predicted spectrum of the WD accretor in black. The WDs are predicted to contribute about half the flux at 1.6\,$\mu$m, but less than 10\% at 5\,$\mu$m. All three sources have strong emission lines. }
    \label{fig:phase_avg_with_wd}
\end{figure}

To identify spectral features and assess the relative importance of the WD, disk, and donor in the spectra, we computed mean spectra by averaging all spectra. We merged data from the G235M and G395M gratings, weighting by inverse variance in the overlap regions.  

In Figure~\ref{fig:phase_avg_with_wd}, we compare these phase-averaged spectra to models of the WD accretors with effective temperatures and radii set to the values inferred from optical SEDs. We calculate the radii of the WDs from their inferred masses (Table~\ref{tab:sample}) using the mass-radius relation from \citet{Bedard2020}. We model the accreting WDs using DB model spectra from \citet{Cukanovaite2021}. These models are completely featureless in the wavelength range covered by our data because they are calculated with a line list that only includes optical transitions. This is not important for this work, since we are primarily interested in the shape and normalization of the WD continuum.

In all cases, the model spectra fall off more steeply at long wavelengths than the data, indicating that the donor and/or disk contribute an increasing fraction of the flux there. Beyond $3\,\mu$m, the observed spectra are nearly flat in $f_{\lambda}$, while the WD models continue to decline steeply. As we will show, the excess emission at long wavelengths is likely dominated by the accretion disk. 

The phase-averaged spectra of all three systems display several prominent emission lines, which are stronger in Gaia14aae and SRGeJ0453 than in ZTFJ1637. All the lines visible in Figure~\ref{fig:phase_avg_with_wd} are He I lines. The spectra and relative line strengths of Gaia14aae and SRGeJ0453 are very similar to one another, while the lines in ZTFJ1637 are weaker.  This could be attributed to the system's longer orbital period and presumably lower mass transfer rate, but it deviates from the general trend seen in the optical, where the equivalent width of AM CVn emission lines is usually largest for long-period systems \citep[e.g][]{Carter2013}.

\subsubsection{Spectra in eclipse}

\begin{figure*}
    \centering
    \includegraphics[width=\textwidth]{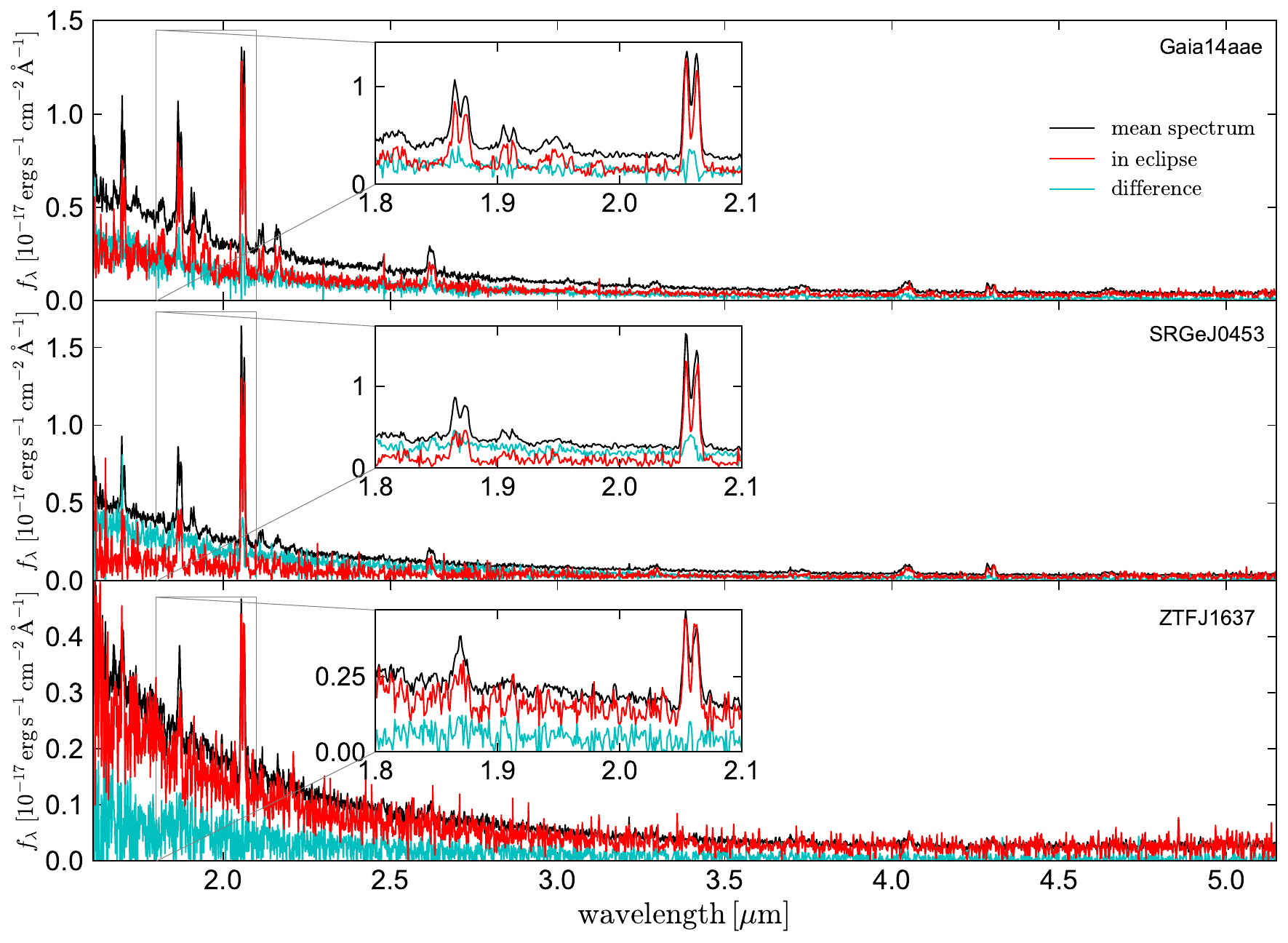}
    \caption{Comparison of phase-averaged spectra (black) to spectra taken during the eclipse of the WD (red). Cyan shows the difference; i.e., the eclipsed component. The in-eclipse spectra are dominated by emission lines, since the donor does not block the whole disk. The eclipsed component, which is dominated by the WD, is nearly featureless. It displays a central spike in several He\,I lines, which may trace emission on or near the accreting WD. }
    \label{fig:eclipse}
\end{figure*}

Figure~\ref{fig:eclipse} compares the phase-averaged spectra to the spectra observed during the primary eclipse. For Gaia14aae and SRGeJ0453, we define the ``in eclipse'' spectrum as the mean of the two exposures taken closest to $\phi = 0$. We only consider the single exposure closest to $\phi = 0$ for ZTFJ1637, since the eclipse duration is less than 1 minute. 

The in-eclipse spectra for Gaia14aae and SRGeJ0453 are $\approx 50\%$ and $\approx 75\%$ fainter than the phase-averaged spectra. Given that both systems have total eclipses of the central WD, this implies that it contributes a larger fraction of the continuum flux in SRGeJ0453, consistent with its higher inferred effective temperature (Figure~\ref{fig:phase_avg_with_wd}). The eclipsed component in both systems is nearly featureless, suggesting it is dominated by the WD, although it also includes contributions from the central regions of the disk. The eclipsed component also features a single-peaked line that resembles the ``central spike'' observed in several AM CVns \citep[e.g.][]{Nather1981, Marsh1999, Kupfer2016, Green2019}. This feature is usually assumed to trace the accreting WD, so it is not surprising to find it in the eclipsed component, but the observed spike in our data might also be due primarily to eclipsed disk material.

\begin{figure}
    \centering
    \includegraphics[width=\columnwidth]{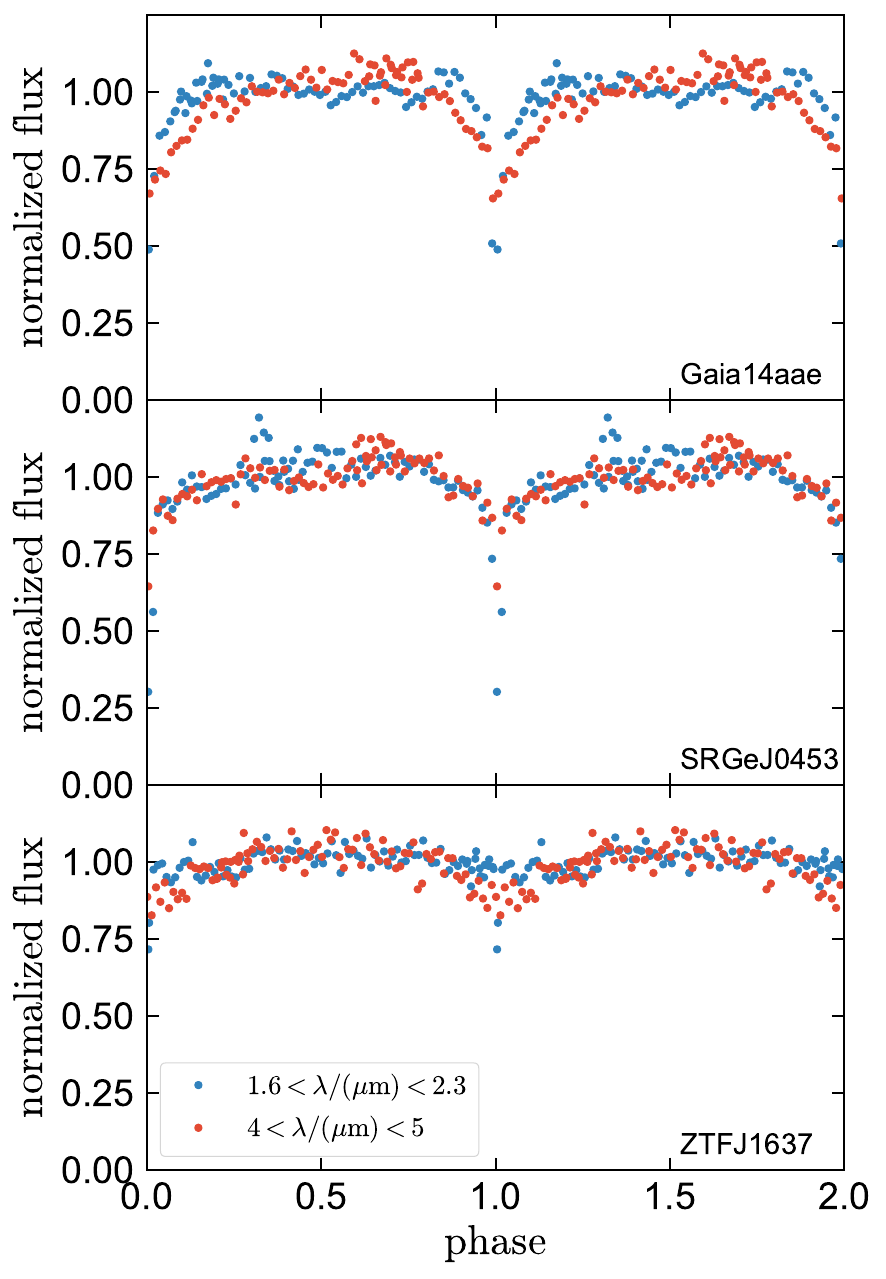}
    \caption{Light curves of the three targets at $1.6-2.3\,{\mu}\rm m$ (blue) and  $4-5\,{\mu}\rm m$ (red). The WD and disk are eclipsed at phase 1. None of the targets show a secondary eclipse at phase 0.5. The primary eclipses are shallower at longer wavelengths, reflecting the fact that the WDs contribute a smaller fraction of the light there. The broad ``wings'' of the eclipse trace the occultation of the disk by the donor. In Gaia14aae and ZTFJ1637, the primary eclipse becomes broader at longer wavelengths. This implies that cooler regions in the outer disk dominate at longer wavelengths, while hotter regions closer to the WD dominate at shorter wavelengths.   }
    \label{fig:lightcurves}
\end{figure}

Figure~\ref{fig:lightcurves} shows light curves of the three systems, with blue and red points showing shorter and longer wavelengths. We calculate light curves by summing all flux in each wavelength range, including both emission lines and continuum, but we find qualitatively similar behavior for both. We calculate phases using the ephemerides in Table~\ref{tab:sample2}, which we find to predict the eclipse times within the $\approx$ 30 second accuracies of our data. A narrow primary eclipse is apparent in the short-wavelength data for all three systems; it is broader and shallower in the long-wavelength data, and barely discernible in ZTFJ1637. The broader disk eclipse is deepest in Gaia14aae, the shortest-period system, perhaps because it has the highest mass transfer rate and thus the brightest disk. None of the three systems show convincing evidence of a secondary eclipse of the donor by the WD and disk at phase 0.5. As we discuss further in Section~\ref{sec:thermal_donors}, this implies that the donor contributes only a small fraction of the total light, even at 4-5\,$\mu$m. 

All three systems show some smooth variability out of eclipse, which is stronger at long wavelengths than at short wavelengths. This could be naturally explained as a result of irradiation of the donor by the WD and disk, such that the ``day'' side of the donor is brighter than the night side. However, given that our data covers only one orbit in each wavelength range and some additional stochastic variability is expected due to the disk, we do not attempt to model the signal quantitatively.

\subsection{Trailed spectra}
\label{sec:trailed}
Figures~\ref{fig:trailed_235m} and~\ref{fig:trailed_395m} show trailed spectra of the three targets in the G235M and G395M gratings. We normalize the data by fitting a third-order polynomial to regions of the phase-averaged spectrum that are free of strong emission lines and dividing all the individual epoch spectra by this mean continuum. 

Most of the lines are double-horned, as is common for disks, but a few are not. The eclipse of the WD is obvious in Gaia14aae and SRGeJ0453, but barely noticeable in ZTFJ1637, which has a grazing eclipse and contains a cooler WD than the other systems. Some changes are detectable in the shape of the He I emission lines over time (see Section~\ref{sec:bright_spot}), but the line centers are not RV variable at a detectable level. The most obviously RV-variable features are two Na I lines in Gaia14aae and SRGeJ0543, which are explored in more detail in Section~\ref{sec:irrad_lines}, and some components of the \spectral{He}{I}{2.058} line, which likely trace the bright spot (Section~\ref{sec:bright_spot}). 

All real emission and absorption lines are resolved across several wavelength pixels. A few narrow vertical stripes are evident in Figures~\ref{fig:trailed_235m} and~\ref{fig:trailed_395m} that are only one or two pixels wide. These are spurious and result from bad pixels that fall directly on the trace in the \texttt{rateints} files (Section~\ref{sec:data_reduction}).

The increasing fraction of the continuum contributed by the disk at long wavelengths is evident from the trailed spectra: in the G235M data, the eclipse is sharp and limited to a few phase bins. A narrow central eclipse is still evident in the G395M data, but it is embedded in a broader, shallower eclipse of the disk.

\subsubsection{Line identification}
\label{sec:lines}
We compared the observed spectra to line lists from \citet{Dhillon1995}, who curated a list of lines likely to be detectable in IR spectra of cataclysmic variables (CVs). Most of the lines they identified that are also present in our data are He\,I lines in emission. There are no clear features detected in absorption. 

Two Na\,I lines are also detected, at wavelengths near 2.21 and 2.34 $\mu$m. Both of these lines are actually doublets, with individual components that are unresolved in our data \citep{Kleinmann1986}. As we discuss in Section~\ref{sec:irrad_lines}, the lines are RV-variable and likely trace the irradiated face of the donor. When we translate the lines' wavelengths to velocities, we adopt a weighted mean vacuum wavelength of 2.2071\,$\mu$m and 2.3373\,$\mu$m for the doublets' rest wavelength, based on the lines' typical ratios from \citet{Kleinmann1986}.

\begin{figure*}
    \centering
    \includegraphics[width=\textwidth]{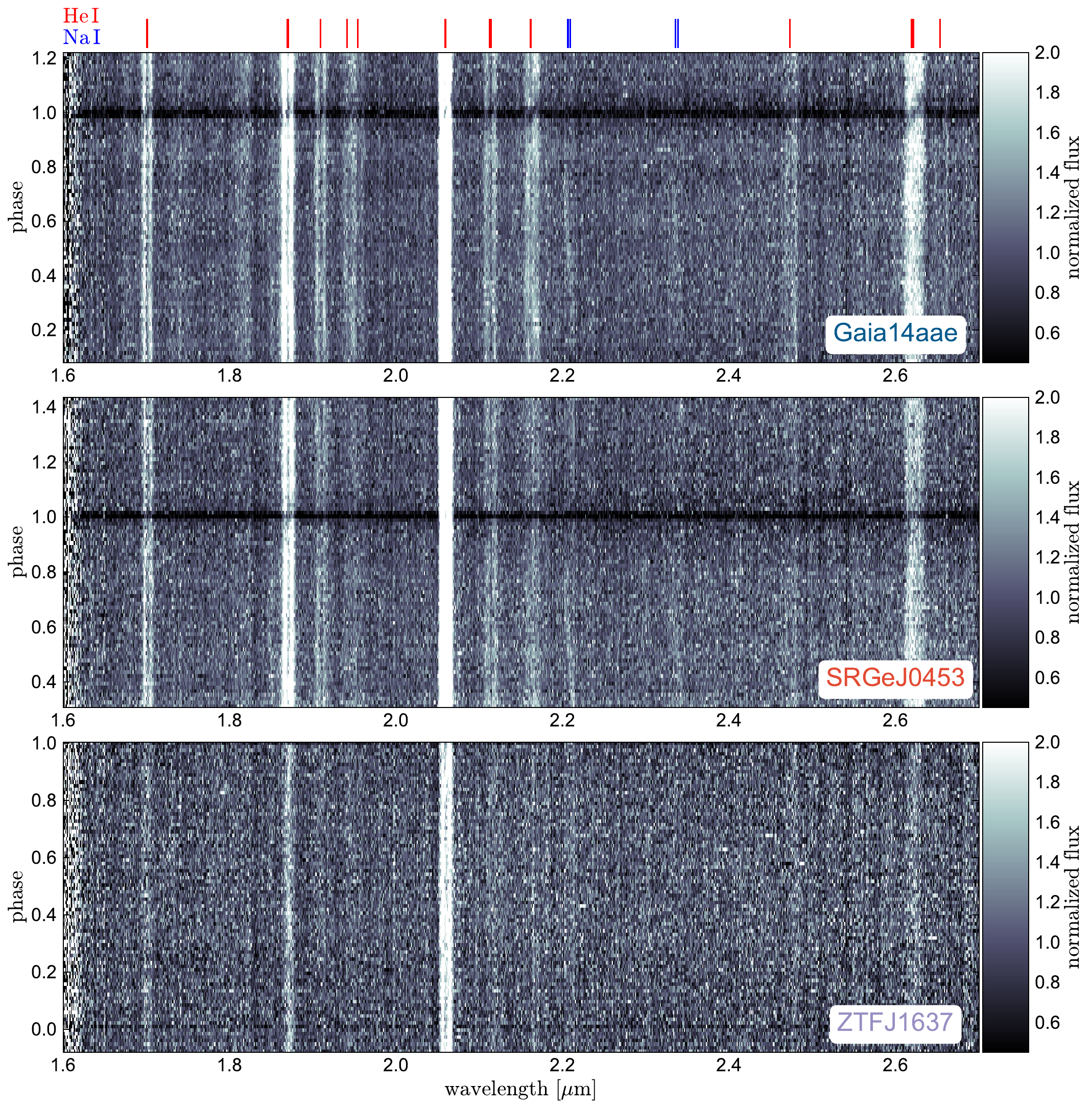}
    \caption{Trailed spectra in the G235M grating. Each 46-second exposure is shown as a single row, with time increasing from the bottom to the top. The eclipse of the WD by the donor occurs at phase 0 and 1. All three objects have spectra dominated by He\,I emission lines, the strongest of which are saturated in the adopted color scale. Most, but not all, of the lines are double-peaked. In Gaia14aae and SRGeJ0453, two Na I lines are also detectable in emission. Unlike the He lines, these are RV-variable and likely trace the irradiated face of the donor (Figure~\ref{fig:trailed_Na}). }
    \label{fig:trailed_235m}
\end{figure*}

\begin{figure*}
    \centering
    \includegraphics[width=\textwidth]{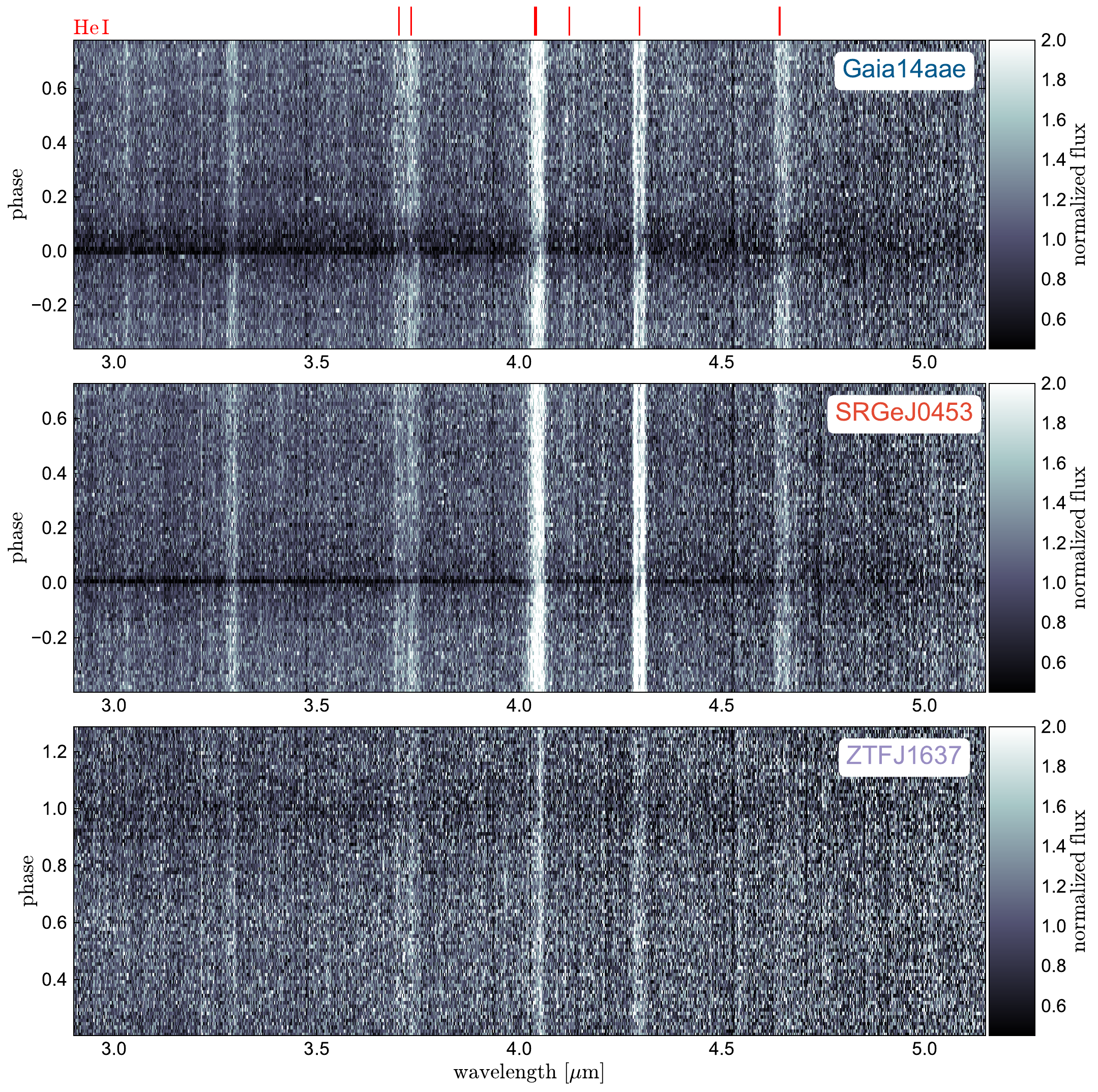}
    \caption{Trailed spectra in the G395M grating. Compared to the shorter-wavelength G235M observations, these data reveal a broader eclipse of the disk, which darkens the continuum for $\approx 15\%$ of the orbit. All the identified lines are due to He\,I. }
    \label{fig:trailed_395m}
\end{figure*}

To identify other lines, we consulted the NIST handbook of basic spectroscopic data \citep{Sansonetti2005}, which lists more than 700 He I transitions between 1.6 and 5.1\,$\mu$m. To identify which of these transitions are likely to produce detectable features in our data, we use ATLAS 12 and SYNTHE \citep{Kurucz1970SAOSR, Kurucz1993sssp} to predict the IR spectrum of a star with a pure He atmosphere and $T_{\rm eff} = 15,000$\,K, which is approximately the temperature of the regions of the disk that dominate the emission. We then identify lines with depth of $\geq 1\%$ relative to the continuum, most of which we find to be detectable in emission in at least one of the observed AM CVns. These lines are marked with red vertical lines in Figures~\ref{fig:trailed_235m} and~\ref{fig:trailed_395m}. There are a few emission lines visible in the data that we could not conclusively identify, the strongest of which is at 3.29\,$\rm \mu m$ in the G395M data. These features could plausibly still be due to He\,I -- the NIST database includes lines at 3.292, 3.293, 3.296, and 3.298 $\mu$m -- but we do not consider the association definitive, because these transitions do not produce strong lines in the Kurucz model spectrum of a He-rich star.

\subsubsection{Irradiation lines tracing the donor}
\label{sec:irrad_lines}
\begin{figure*}
    \centering
    \includegraphics[width=\textwidth]{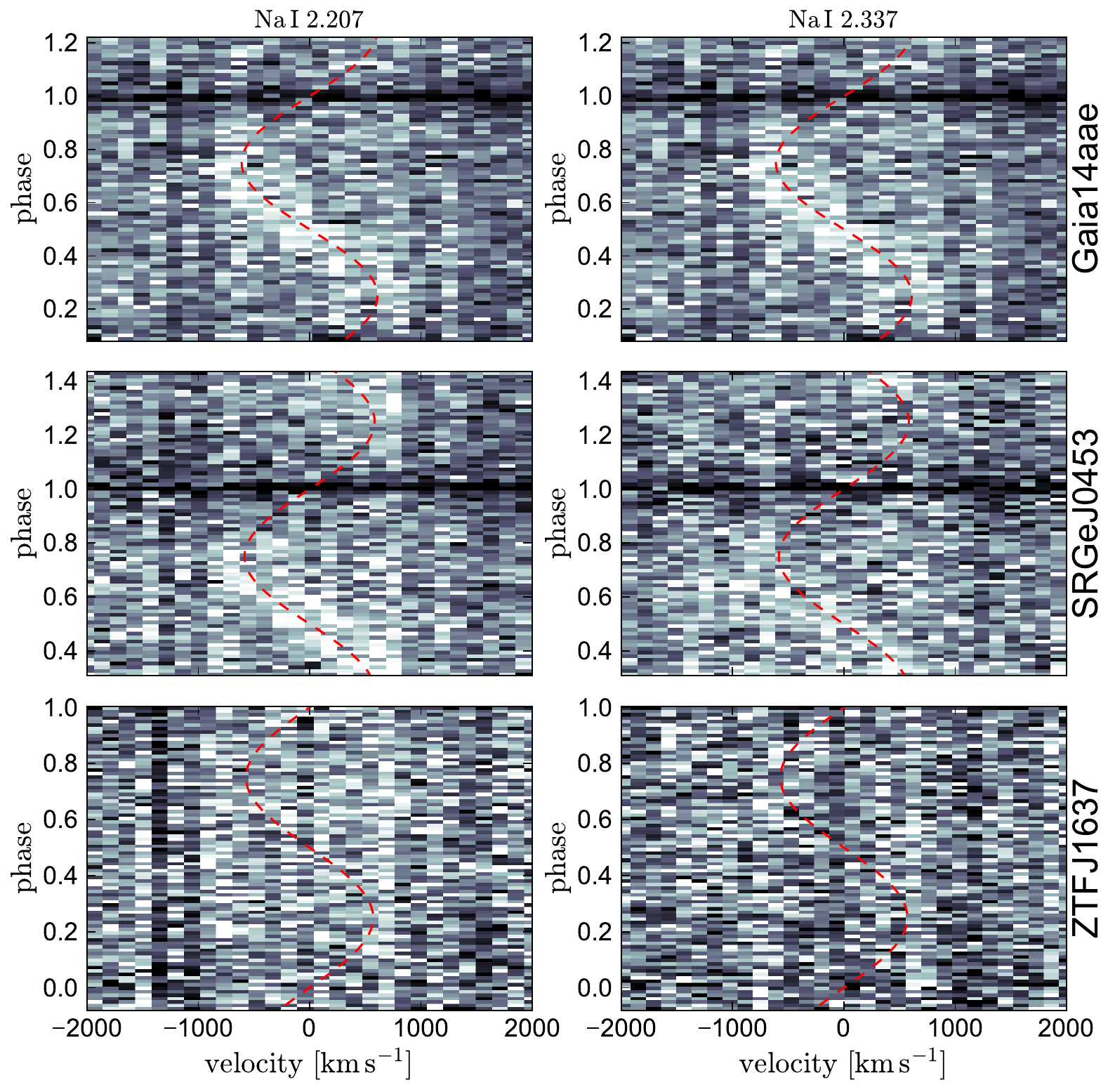}
    \caption{Trailed spectra of two Na\,I lines, each of which is an unresolved doublet. Dashed red curves show the predicted RV of the donor. In Gaia14aae and SRGeJ0453, emission lines are visible over more than half the orbit and trace these curves. We interpret these lines as originating on the irradiated ``day'' side of the tidally-locked donors. The lines are only visible when the donor is behind the WD and its irradiated side faces Earth. The lines are not robustly detected in ZTFJ1637.}
    \label{fig:trailed_Na}
\end{figure*}

Figure~\ref{fig:trailed_Na} shows the two sets of RV-variable Na I doublets in all three systems as a function of velocity and phase. We overplot the predicted RV curve of the donor, which we calculate from the assumed masses and inclinations in Table~\ref{tab:sample}, assuming circular orbits. The observed emission lines in Gaia14aae and SRGeJ0453 are consistent with these predictions and are detectable for $\approx 60\%$ of the orbit. They are not detected in ZTFJ1637. 

In Gaia14aae and SRGeJ0453, the observed emission lines are strongest near phase 0.5. They are completely absent near phase 1, when the donor is in front of the WD. This behavior is naturally explained if the emission arises on the irradiated side of the donor that faces the WD. At phase 0.5, the strongest line rises about 30\% above the continuum, with a maximum equivalent width of $\approx 10$\,\AA. Since the amplitude of the emission lines relative to the donor's continuum is uncertain, the continuum flux ratio is not well constrained. Nevertheless, the detection of the RV-variable Na\,I lines in Gaia14aae and SRGeJ0453 is to our knowledge the first clear detection of the donor in a long-period AM CVn.

Na enhancement is often observed in CVs with evolved donors \citep[e.g.][]{Thorstensen2002, El-Badry2021, Yamaguchi2023}, where it has been explained as a result of $^{23}$Na production in the NeNa cycle at the end of the donors' main-sequence burning. Such CVs are one proposed progenitor for AM CVns \citep[e.g.][]{Belloni2023}. Given that Na is the only element we detect from the donors in our data, it is tempting to attribute the lines in Figure~\ref{fig:trailed_Na} to Na enhancement and interpret this as evidence for the evolved CV channel. However, we have not actually measured an Na abundance, and establishing Na enhancement will required robust spectral models for the donors, including the effects of irradiation. We defer such investigation to future work. 

\begin{figure*}[!t]
    \centering
    \includegraphics[width=\textwidth]{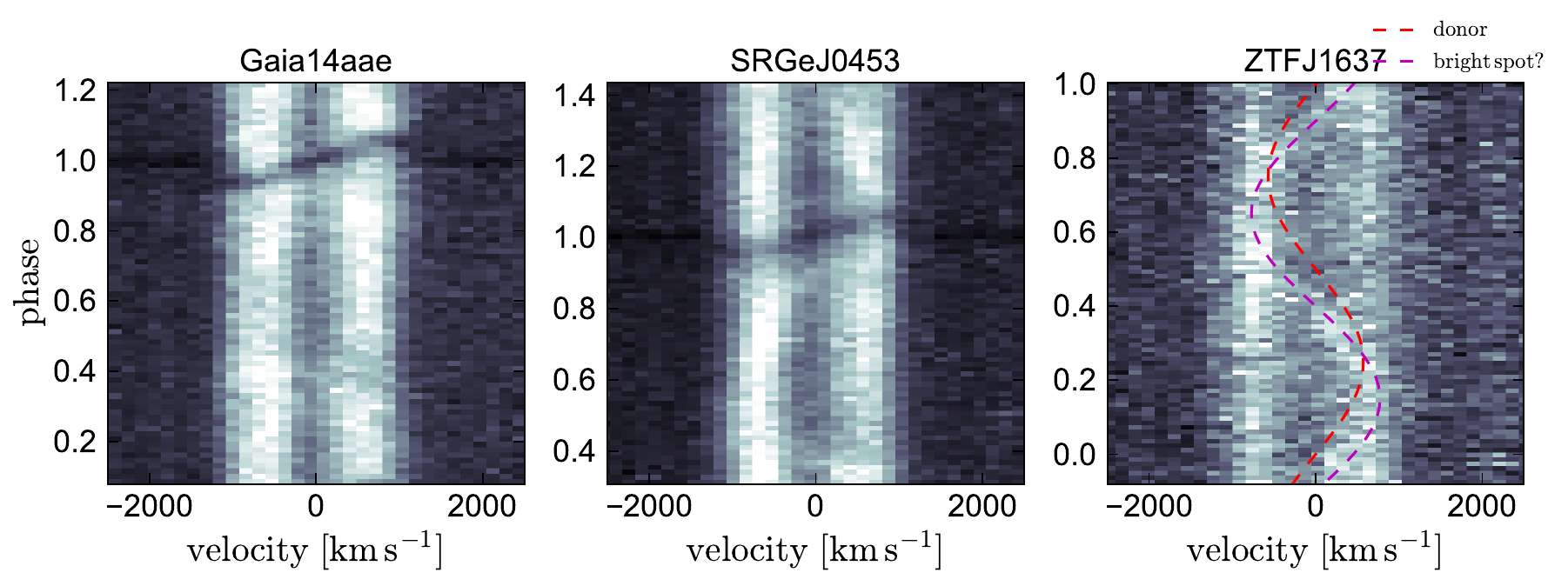}
    \caption{Trailed spectra of the He\,I 2.058 line. In Gaia14aae and SRGeJ0453, the eclipse is seen to move across the disk, with the blueshifted wing eclipsed before the WD eclipse, and the redshifted wing eclipsed after. The disk eclipse extends over a period $\sim$5 times longer than the eclipse of the WD. The eclipse is less visible in ZTFJ1637. In ZTFJ1637, a clear sinusoidal ``S wave'' is visible. Red line shows the predicted RV of the donor, which is offset in phase from the data and has a smaller RV amplitude. The observed sinusoid likely traces the ``bright spot'', where the accretion stream impacts the disk. Dashed magenta line shows a sinusoid offset in phase by 0.1 from the donor, with a 35\% larger RV amplitude. This provides a good fit to the RVs of the bright spot, yielding constraints on the mass ratio and accretion disk radius (Figure~\ref{fig:bright_spot}). }
    \label{fig:trailed_HeI}
\end{figure*}

\subsubsection{Variability in the He\,I lines}
\label{sec:bright_spot}

Figure~\ref{fig:trailed_HeI} shows spectrograms of the \spectral{He}{I}{2.058} line, which is the strongest line present in the data. All three systems show phase-dependent variability in the profile of the line, with coherent changes in the amplitude of the red and blue peaks. In addition, all three systems show an RV-variable emission component that crosses the center of the line near phase 0.4 and 0.9. This component is most obvious in ZTFJ1637 (right panel), where it has a characteristic ``S wave'' shape and is easily recognizable because other variability in the line is weak.

\subsubsection{Modeling the bright spot in ZTFJ1637}
\label{sec:bright_spot}

The red line in Figure~\ref{fig:trailed_HeI} shows the expected orbit of the donor. Unlike the Na\,I lines in Figure~\ref{fig:trailed_Na}, the S wave in ZTFJ1637 is obviously offset in phase from the donor. It also has a larger RV semi-amplitude than predicted for the system's inferred component masses and inclination.  Unlike the Na\,I lines, which trace the donor, the emission in He\,I likely traces the ``bright spot'', where the accretion stream collides with the outer edge of the disk.

Fitting the bright spot RVs in the \spectral{He}{I}{2.058} line of ZTFJ1637 with a sinusoidal model, we infer $K_{\rm bright\,spot} = 779\pm 21\,{\rm km\,s^{-1}}$, which is 35\% larger than the predicted RV semi-amplitude of the donor. We find a phase offset of $\Delta \phi_{\rm bright\,spot} = 0.90\pm 0.02$ (or equivalently, $-0.10\pm 0.02$) relative to the donor. This best-fitting sinusoid is plotted in magenta in Figure~\ref{fig:trailed_HeI}.

The RV semi-amplitude and phase offset of the bright spot constrain the  trajectory of the accretion stream, and thus, the mass ratio and the size of the accretion disk. Following \citet{Lubow1975}, we integrate the equations of motion for a test particle released from the L1 Lagrange point in the Roche potential until it intersects the disk \citep[see e.g.,][]{Roelofs2006, Kupfer2016}. The bright spot likely traces this intersection point. It arises from the collision of material in the accretion stream with material in the outer disk, so we predict its velocity as the vector average of the stream velocity and the Keplerian velocity at that position.

\begin{figure*}[!t]
    \centering
    \includegraphics[width=\textwidth]{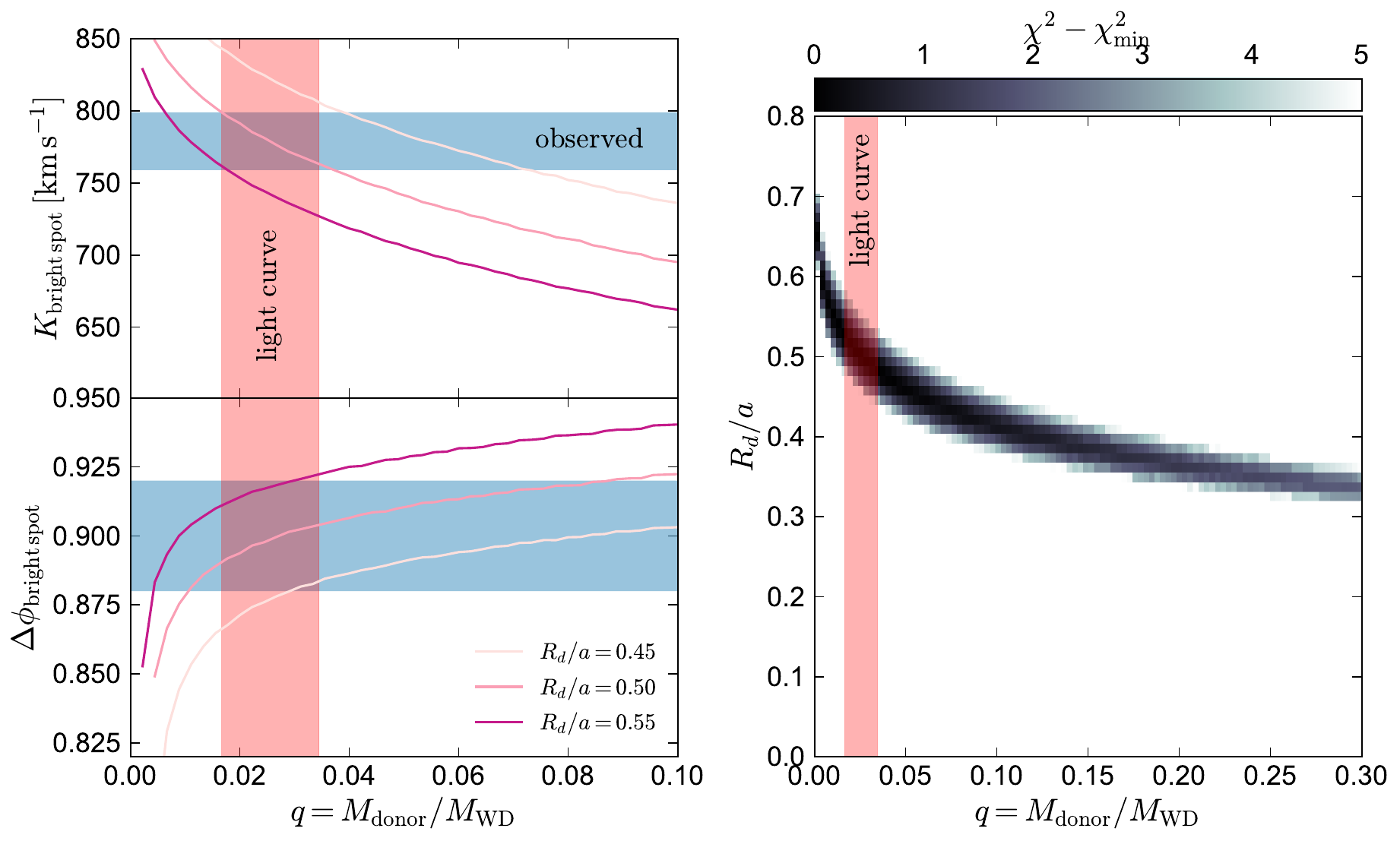}
    \caption{Left: bright spot RV semi-amplitude (top) and phase offset relative to the donor (bottom) in ZTFJ1637. Blue shaded region shows the observed value; red shading shows mass ratio constraints from optical light curves. Colored lines show results of integrating stream trajectories until they collide with the disk for different disk outer radii.  Right: joint constraints on the mass ratio and disk size from the observed bright spot RV semi-amplitude and phase offset. When the light curve constraint on the system's mass ratio is enforced, the bright spot trajectory implies $R_d/a = 0.50\pm 0.03$.  }
    \label{fig:bright_spot}
\end{figure*}

Figure~\ref{fig:bright_spot} shows the results of these calculations for ZTFJ1637. In the left panels,  we consider three possible values for the outer disk radius, $R_{d}$, and integrate stream trajectories for a range of mass ratios. If both $R_d$ and $q$ are left free, the data are consistent with a rather broad range of mass ratios, with lower values of $q$ corresponding to more extended disks. However, if we fix $R_d$ to the radius of the largest stable streamline, $R_d \approx 0.48a$ \citep{Paczynski1977}, the constraint tightens to $0.02 < q < 0.04$, which is fully consistent with the constraints from light curve modeling (Figure~\ref{tab:sample}). Alternatively, if we set $q=0.026$ as inferred from light curve fitting, then the data imply $R_d/a = 0.50\pm 0.02$. We show joint constraints on $q$ and $R_d/a$ when both parameters are left free in the right panel of Figure~\ref{fig:bright_spot}. Marginalized constraints on $q$ and $R_d/a$ from the bright spot alone are weak, because the two parameters are highly covariant, but the constraints on $q$ from light curve modeling break this degeneracy. 

Constraints from the bright spot RVs in ZTFJ1637 are thus broadly consistent with the light curve solution from \citetalias{vanRoestel2022}. We do not attempt similar fits for Gaia14aae or SRGeJ0453, because the morphology of the He\,I lines is more complex and the bright spot's RV variability cannot be described by a single sinusoid. Below, we use Doppler tomography to interpret the line profiles of these systems.

\subsubsection{Doppler Tomography}
\label{sec:dopto}

Doppler tomograms are two-dimensional maps that reconstruct the distribution of line emission in velocity space. They show where different components contribute in RV and orbital phase, allowing emission features from the accretion disk, bright spot, and donor star to be disentangled in velocity coordinates \citep[e.g.][]{Marsh1988, 2001marsh}. In such maps, the radial coordinate represents the projected RV, and the azimuthal coordinate corresponds to orbital phase, so that each pixel encodes the velocity (rather than the spatial) coordinates of the emitting gas. We use the IDL-based code \texttt{doptomog} \citep{2015kotze} to generate Doppler tomograms for several emission lines.

\begin{figure*}
    \centering
    \includegraphics[width=\textwidth]{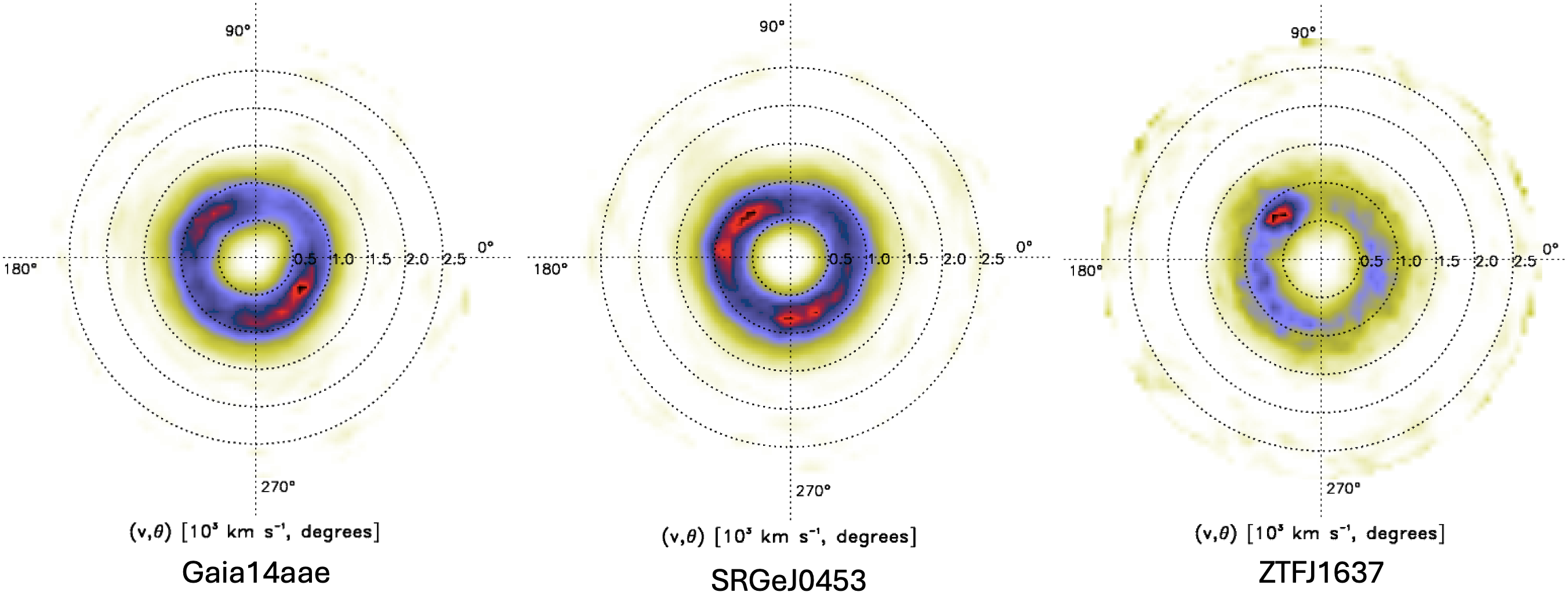}
    \caption{Doppler tomograms of the He I 2.058 $\mu$m emission line. The location of the donor star is at the 90$^\circ$ phase, and the WD is in the center. A clear disk is revealed in all systems. Gaia14aae and SRGeJ0453 show evidence of two bright spots, located at nearly opposite phases from each other. ZTFJ1637 yields the only tomogram resembling that of most CVs and AM CVns, featuring a disk and single bright spot just ahead of the donor star in orbital phase.}
    \label{fig:doppler_2.058}
\end{figure*}

\begin{figure*}
    \centering
    \includegraphics[width=0.9\textwidth]{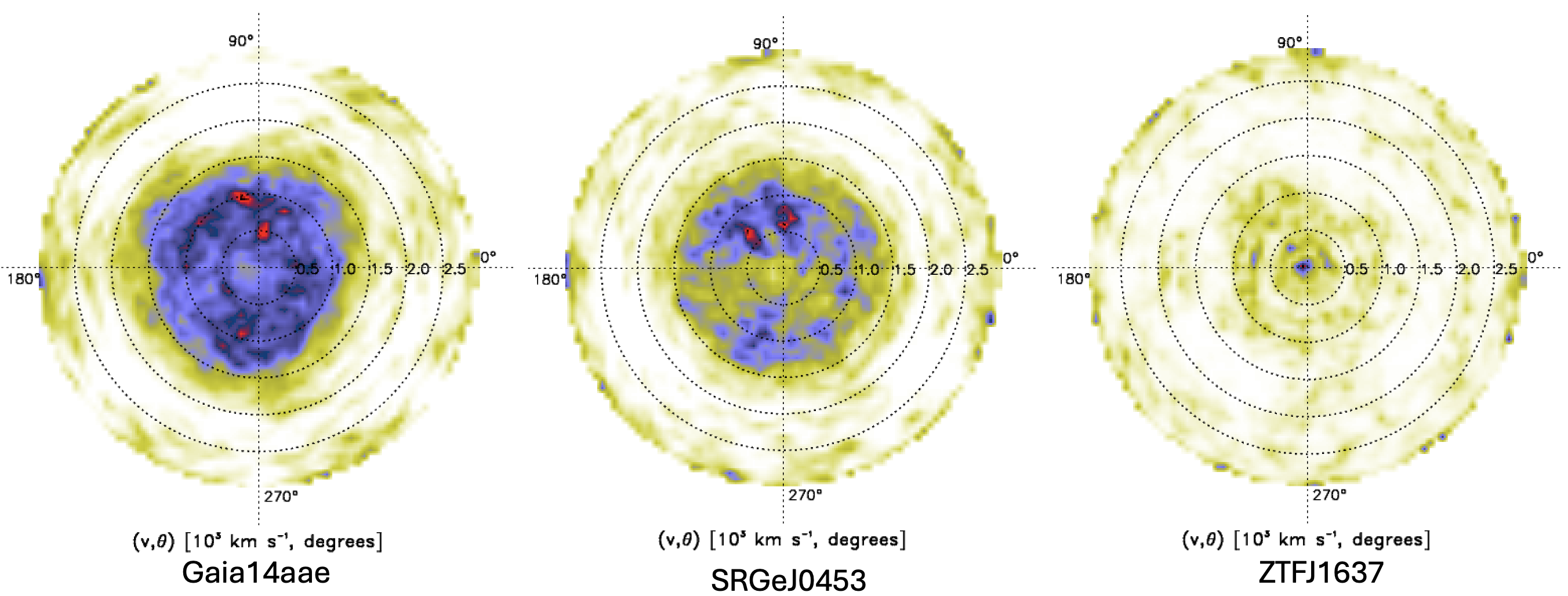}
    \caption{Doppler tomograms of the He I 2.621 $\mu$m emission line. Unlike the He I 2.058 $\mu$m line, no disk is seen; this reflects the fact that the line is not double-peaked. In both Gaia14aae and SRGeJ0453, the line appears to trace velocities from zero to $\approx$1000 km s$^{-1}$, with marginal evidence for increased emission near the donor star at 90$^\circ$ phase. The line is only marginally detected in ZTFJ1637.  }
    \label{fig:doppler_2.621}
\end{figure*}

\begin{figure*}
    \centering
    \includegraphics[width=0.9\textwidth]{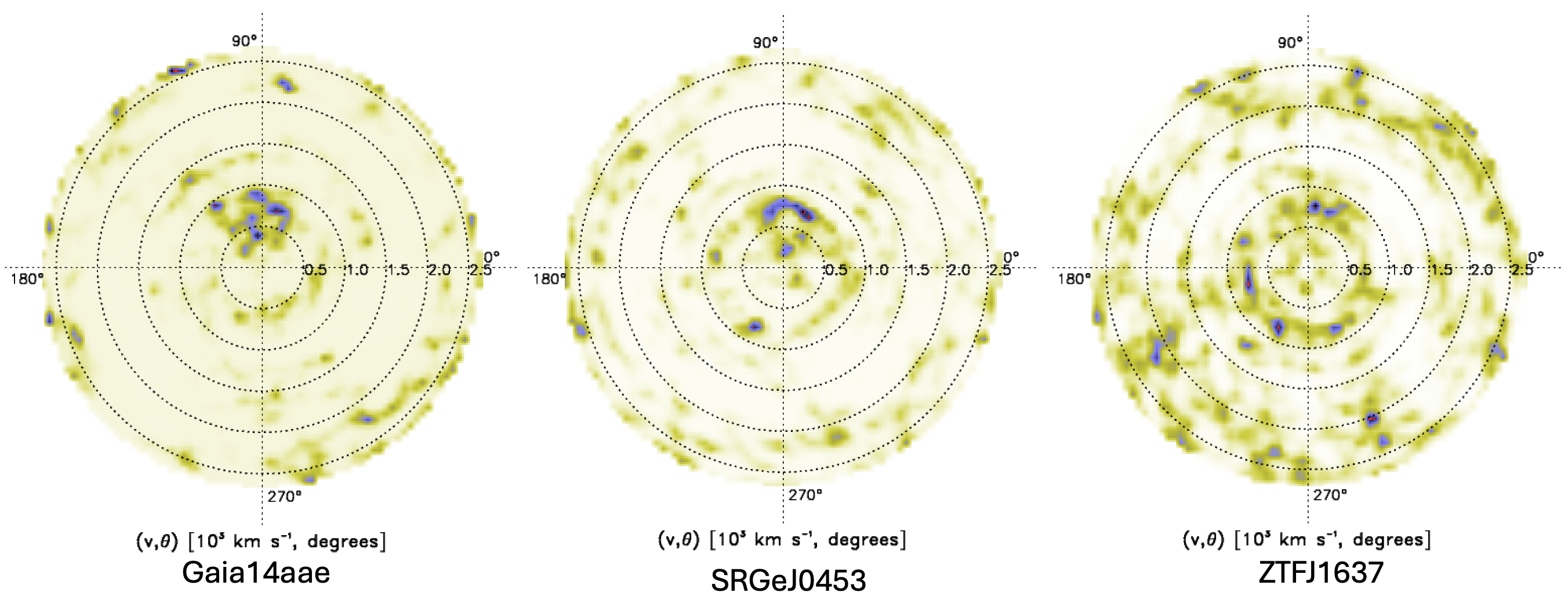}
    \caption{Doppler tomograms of the Na I 2.207 $\mu$m emission line. In both Gaia14aae and SRGeJ0453, the line traces a region near the donor star (90$^\circ$ phase). This line is only marginally detected in ZTFJ1637.}
    \label{fig:doppler_2.207}
\end{figure*}

Figures \ref{fig:doppler_2.058}, \ref{fig:doppler_2.621}, and \ref{fig:doppler_2.207} show tomograms of the He I 2.058 $\mu$m, He I 2.621 $\mu$m, and Na I 2.207 $\mu$m emission lines, respectively. Doppler tomograms of the He I 2.058 $\mu$m line show a disk in all three systems with a clear bright spot just ahead of the donor star, where the stream intersects the disk. However, Gaia14aae and SRGeJ0453 have two bright spots, possibly indicative of the second impact spot of the stream on the other side of the disk \citep[e.g.][]{Hessman1999, Tappert2003, Zharikov2013, Kupfer2013, Longa-Pena2015, Neustroev2016, Kupfer2016}. Some evidence for this has been seen in the optical in the case of Gaia14aae \citep[e.g.][]{Green2019}, though here, the two spots appear to be of equal intensity. 

Figure \ref{fig:doppler_2.621} does not reveal a clear disk in any system; indeed, the He I 2.621 $\mu$m line is not doubled in any of the trailed spectra. This line is actually a pair of unresolved lines, with two transitions separated by $\approx 200\,{\rm km\,s^{-1}}$, but since this separation is small compared to the velocity separation of the two peaks present in other lines, the presence of two transitions is unlikely to be the main reason for the line's different profile and lack of double-peaked emission. We conjecture that \spectral{He}{I}{2.619}, which is a triplet, has a higher optical depth than the singlet \spectral{He}{I}{2.058} line \citep[e.g.][]{Honeycutt2013}, making it more sensitive to low-density material and thus a less reliable kinematic tracer of the disk.

Figure \ref{fig:doppler_2.207} shows the Doppler ``image'' of the donor stars in the Na I 2.207 doublet. As expected, there is increased emission in Gaia14aae and SRGeJ0453 at phase 90$^\circ$ and velocity $\approx 600\,{\rm km\,s^{-1}}$, which is the expected location of the donor. The signal-to-noise ratio is somewhat lower than might be expected from Figure~\ref{fig:trailed_Na}, because the line is only visible for $\sim 60\%$ of the orbit.

\subsection{Eclipse of emission lines}
\label{sec:eclipse_mapping}

In Gaia14aae and SRGeJ0453, the eclipse can be observed to move across the double-peaked emission lines in velocity space (Figure~\ref{fig:trailed_HeI}), with the blueshifted edge of the line eclipsed before the start of the eclipse of the WD (visible in the continuum) and the redshifted side of the line eclipsed after the WD eclipse has ended. While the WD is eclipsed for only $\lesssim 2$ minutes in both systems, the disk eclipse takes 8-10 minutes. Hints of velocity-dependent eclipse times were evident in optical spectra of Gaia14aae presented by \citet{Green2019}, but the eclipse was less well-resolved, with only two phase bins covering the phase interval that is spread over 12 bins here. As we explore below, the shape of the disk eclipse profile constrains the disk geometry.

No comparable disk eclipse is visible in the emission lines of ZTFJ1637, where the eclipse is grazing and the lines are weaker. However, an extended eclipse of the disk continuum can still be observed in Figure~\ref{fig:trailed_395m}, where it manifests as a broad shadow that moves across the spectrogram near phase 1.

\subsubsection{Eclipse mapping}
\label{sec:interpretation}

The shape of the line eclipses contains information about the binary orbit and the morphology of the disk \citep[e.g.][]{Young1980, Horne1995}. To interpret the observed emission line eclipses, we use simple simulations of the eclipse of an optically thick disk by an opaque donor. We model the disk as a sum of 1000 concentric annuli, extending from $r=r_{\rm min}$ to $r = r_{\rm max}$. We divide each annulus into 1000 azimuthal sections, each with a line-of-sight velocity $v_{{\rm los}}=\sqrt{GM_{{\rm WD}}/r}\sin i\sin\varphi$ relative to the observer. Here $i$ is the inclination and $\varphi$ the azimuthal angle, with $\varphi=0$ pointing toward the observer. 

We follow \citet{Horne1986} in modeling the emission from each patch in the optically thick limit, accounting for both thermal broadening and shear. We assume that hydrostatic equilibrium results in a thermal velocity of $v_{\rm th} \sim (H/R)v_\phi$, where $H$ and $R$ represent the vertical and horizontal scale heights of the disk. We assume $H/R=0.04$. We assume the line source function in the emission layer is a power law, $S_L(r) \propto r^{-s}$. We adopt $s=1$ as a fiducial scaling following \citet{Horne1986} but also experiment with other values. We set the component masses and radii to match the observed values for Gaia14aae (Table~\ref{tab:sample}).


During the eclipse, material in the disk that is geometrically occulted by the donor -- whose radius we take from Table~\ref{tab:sample} -- does not contribute to the total flux. We similarly remove light from the WD when it is occulted, modeling both the WD and donor as spheres.  Motivated by \citet{Paczynski1977}, we set $r_{\rm max} = 0.48a \approx 0.21\,R_{\odot}$, which is the radius of the largest stable streamline. We vary $r_{\rm min}$, the inner edge of the disk. 

\begin{figure*}[!t]
    \centering
    \includegraphics[width=\textwidth]{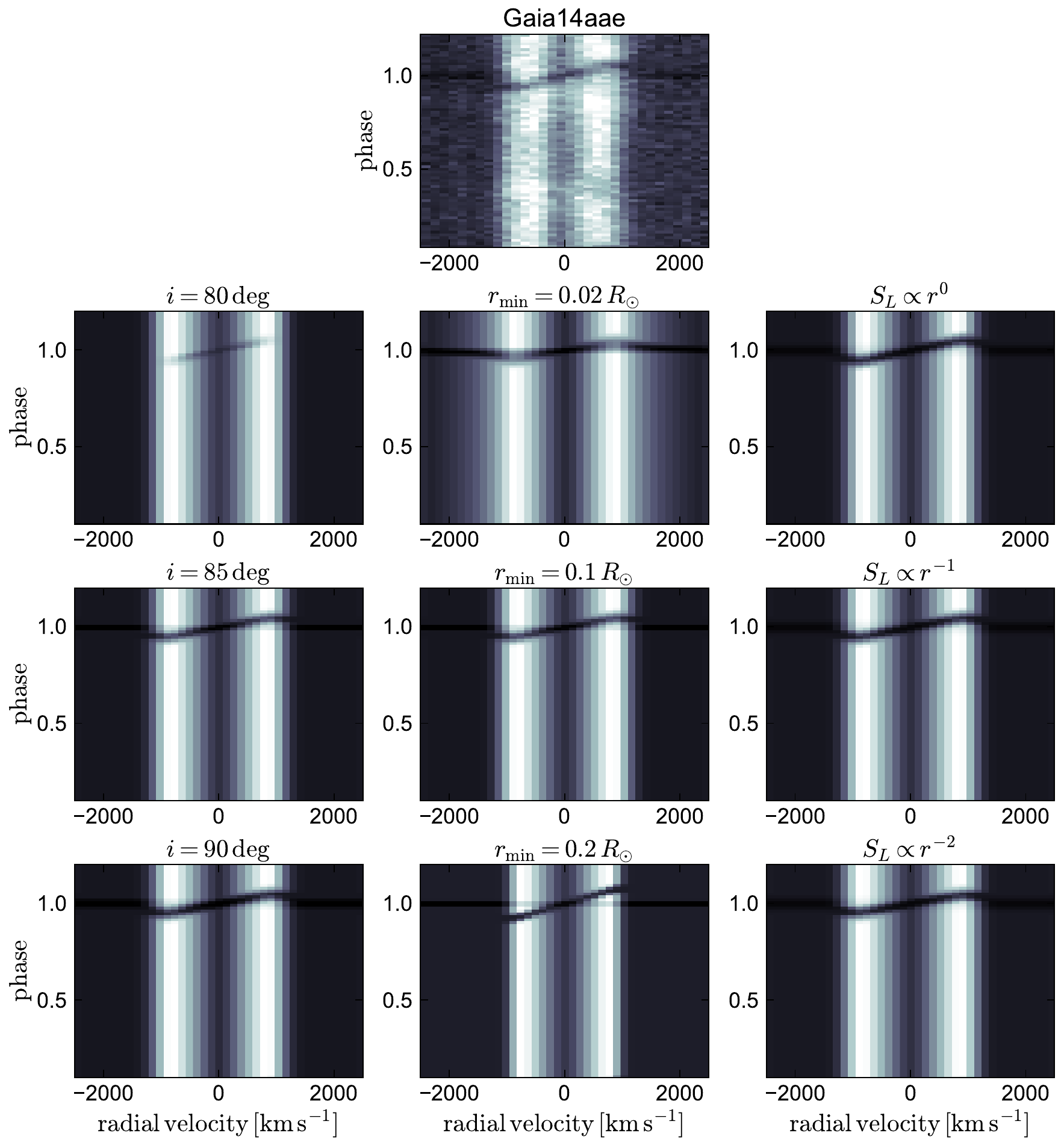}
    \caption{Disk line eclipses. Top panel shows the \spectral{He}{I}{2.058} line in Gaia14aae. Next three rows show simulated spectrograms. We model line emission from an optically-thick disk occulted by the donor at phase 1 (Section~\ref{sec:interpretation}). We vary the inclination (left column), inner disk radius (middle column) and radial dependence of the source function (right column); we adopt fiducial values of $i = 85$ deg, $r_{\rm min} = 0.1\,R_{\odot}$, and $S_L \propto r^{-1}$ in each column.  Varying the inclination primarily changes the depth of the disk eclipse. Varying the inner disk radius changes the disk emission line profile and the duration of the line eclipse. Varying the source function's radial dependence has only modest effects on the line profile and eclipse shape. Comparing to the observed disk eclipses of Gaia14aae and SRGeJ0453 (Figure~\ref{fig:trailed_HeI}), the line eclipse shapes are best-matched by models with $r_{\rm min} \approx 0.1\,R_{\odot}$. A significantly smaller $r_{\rm min}$ predicts too much flux at large velocity and an insufficiently curved line eclipse.  }
    \label{fig:sim_spectrogram}
\end{figure*}

Figure~\ref{fig:sim_spectrogram} shows predicted spectrograms for a few choices of inclination, $r_{\rm min}$, and disk emissivity profile. Since different emissivity profiles $S_L$ lead to different total line flux, we re-scale the continuum flux in each panel to a fixed fraction of the maximum line flux. We compare these model predictions to the observed \spectral{He}{I}{2.058} line in Gaia14aae, which is shown in the top row. 

All the simulated profiles display a ``tilde''-shaped eclipse that moves across the emission line. The peaks of the emission line are dominated by material near $r=r_{\rm max}$, where the disk has the largest area. The line wings are dominated by light at small radii, where the orbital velocity is highest. Due to projection effects, the emission  at small velocities traces material at all radii near $\varphi = 0$ or $\varphi = \pi$, where orbital motion is perpendicular to the line of sight. The outer edge of the disk is eclipsed first, so the peaks of the emission line profile are eclipsed before the wings or the center.

The clearest conclusion we can draw is that $r_{\rm min}$ must be considerably larger than the radius of the WD ($\approx 0.01\,R_{\odot}$), with a value of $r_{\rm min}$ between $0.05\,R_{\odot}$ and $0.1\,R_{\odot}$ fitting the data best. A smaller value of $r_{\rm min}$ both produces emission lines that are wider than observed -- since the Keplerian velocity increases toward the WD surface -- and produces an eclipse velocity map inconsistent with observations, with the line wings being eclipsed later. 

The minimum radius of emitting material in the disk is directly related to the maximum velocity of emission line wings, $v_{\rm max}$. If the emission lines are intrinsically narrow and are broadened by rotation of gas in an edge-on Keplerian disk, the minimum disk radius is:

\begin{equation}
    \label{eq:rmin}
    r_{{\rm min}}=0.067\,R_{\odot}\left(\frac{v_{{\rm max}}}{1500\,{\rm km\,s^{-1}}}\right)^{-2}\left(\frac{M_{{\rm WD}}}{0.8\,M_{\odot}}\right).
\end{equation}
Equation~\ref{eq:rmin} represents a conservative lower limit on $r_{\rm min}$, since emission line wings are further broadened by non-coherent scattering, turbulence, and instrumental broadening. 

\begin{figure*}
    \centering
    \includegraphics[width=\textwidth]{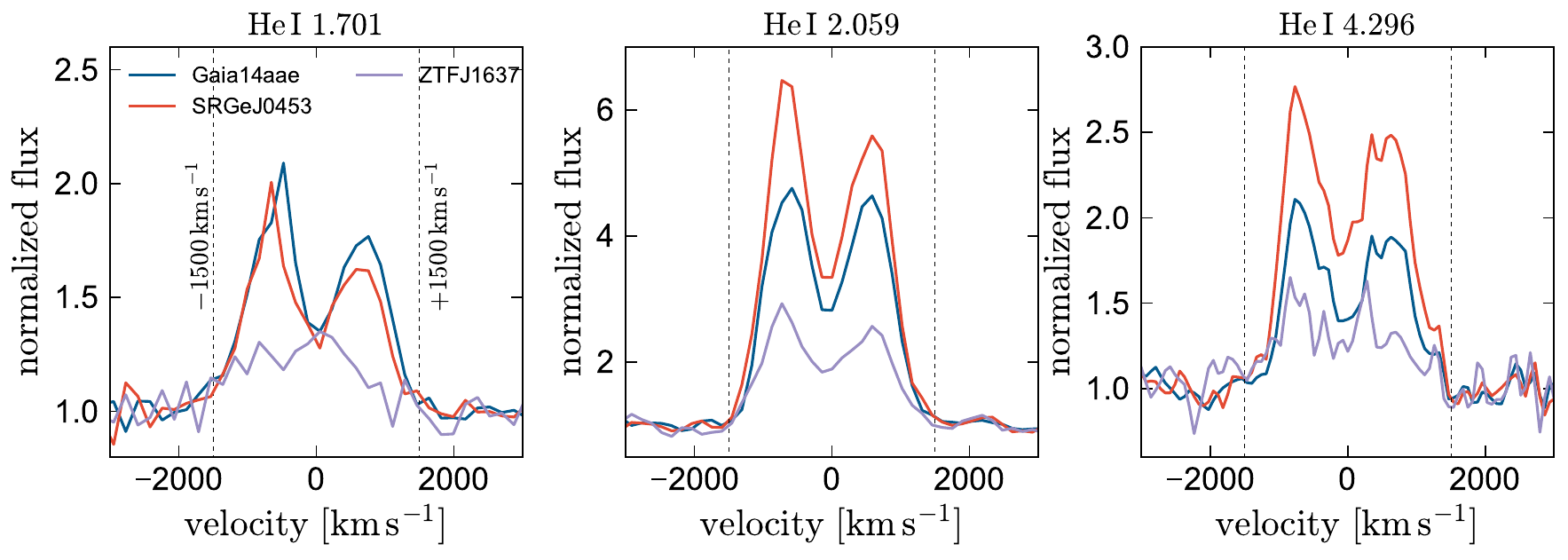}
    \caption{Phase-averaged profiles of three strong and unblended He\,I lines. Gaia14aae and SRGeJ0453 have similar double-peaked profiles in all lines, while ZTFJ1637 has weaker lines that in some cases are single-peaked. The line shoulders extend to $\pm 1500\,{\rm km\,s^{-1}}$ (dashed vertical lines). The lack of emission at higher velocities implies a truncation in the inner disk, with little or no emission inside $r \lesssim 0.07\,R_\odot$ (Equation~\ref{eq:rmin}).  }
    \label{fig:profiles}
\end{figure*}

Figure~\ref{fig:profiles} compares the line profiles of three He I emission lines in the three targets. In all cases, emission is limited to within $\pm 1500\,{\rm km\,s^{-1}}$ of line center. This is consistent with an inner disk radius of $r_{\rm min} \approx 0.07\,R_\odot$, and much slower than the maximum expected velocity of $\approx 4000\,{\rm km\,s^{-1}}$ if the disk extends to the surface of the WD. Given the difficulty of uniquely identifying the continuum when producing normalized emission line profiles, it remains possible that some very faint emission is present at high velocities \citep[e.g.][]{Smak1981}. However, our analysis rules out an emissivity profile that continues to rise at small radii, requiring of a break in the emissivity near $r_{\rm min}$. 

One might speculate that the inner regions of the disk are hotter, and do not emit in He\,I because the gas is ionized. This, however, is unlikely to be the case, because we do not observe broad He\,II emission. He\,II is detected in all three systems in the optical, but only as a narrow, single-peaked line tracing the WD and boundary layer \citep[e.g.][]{Green2019}. The only He\,II line covered by our data is at 1.864\,$\mu$m, but this line is weak and blended with \spectral{He}{I}{1.869}. Other AM CVns have been observed to exhibit He\,II lines that are significantly broader than their He\,I lines \citep[][]{Marsh1999, Roelofs2009}.

\subsection{Evidence for magnetic disk truncation?}
\label{sec:magnetic}
The emission line profiles of all three AM CVns in our sample suggest that the disks are truncated at $r_{\rm min} = (0.05-0.10)\,R_{\odot}$. A possible explanation is that the disks are truncated by the WD's magnetic field, which funnels the accretion disk into a stream that channels inflowing gas to the accreting WD's magnetic poles. In the case of spherical accretion, inflow will be truncated near the Alfv\'{e}n radius, 

\begin{equation}
    \label{eq:rA}
    r_{{\rm A}}=\left[\frac{\mu_{{\rm WD}}^{4}}{2GM_{{\rm WD}}\dot{M}}\right]^{1/7},
\end{equation}
where $\mu_{\rm WD}=B_{\rm WD} R_{\rm WD}^3$ is the WD's magnetic moment, $B_{\rm WD}$ is its equatorial surface magnetic field, $R_{\rm WD}$  is its radius, and $\dot{M}$ is the mass transfer rate. Equation~\ref{eq:rA} represents the condition for the WD's magnetic pressure to balance ram pressure, such that at $r \lesssim r_{\rm A}$, inflowing gas flows along magnetic field lines and is deposited near the poles.

If the accretion flow forms a disk, magnetic truncation is expected to occur at a radius $r_{\rm min}\approx 0.5 r_A$ \citep[e.g.][]{Ghosh1979}. This leads to an expression for the equatorial surface field of the WD, $B_{\rm WD}$, in terms of the inner disk radius:

\begin{equation}
\begin{split}
    B_{{\rm WD}} = 86\,{\rm kG}
    \left(\frac{r_{{\rm min}}}{0.07\,R_{\odot}}\right)^{7/4}
    \left(\frac{M_{{\rm WD}}}{0.8\,M_{\odot}}\right)^{1/4} \\
    \times
    \left(\frac{\dot{M}}{10^{-11}M_{\odot}\,{\rm yr^{-1}}}\right)^{1/2}
    \left(\frac{R_{{\rm WD}}}{0.01\,R_{\odot}}\right)^{-3}
\end{split}
\label{eq:Beq}
\end{equation}

The mass transfer rates of AM CVn binaries are somewhat uncertain. Evolutionary models predict $\dot{M} = 10^{-12} - 10^{-11}\,M_{\odot}\,\rm yr^{-1}$ at orbital periods of 50-65 min \citep[e.g.][]{Deloye2005, Bildsten2006, Wong2021}. However, inferred temperatures of observed long-period AM CVns are higher than predicted for such low $\dot{M}$ values \citep[e.g.][]{Ramsay2018, vanRoestel2022, Macrie2024}, suggesting higher mass transfer rates of order $10^{-10.5}\,M_{\odot}\,\rm yr^{-1}$ at periods near 50 min. It is unclear whether this reflects an additional source of angular momentum loss in AM CVns \citep[such as magnetic braking; e.g.][]{Farmer2010, Sarkar2023} or systematically biased WD temperature measurements. In any case, the dependence on $\dot{M}$ in Equation~\ref{eq:Beq} is relatively weak, such that $r_{\rm min} \approx 0.07\,R_{\odot}$ implies $30 \lesssim B_{\rm WD}/{\rm kG} \lesssim 300$ for $-12 < \log\left[\dot{M}/\left(M_{\odot}{\rm yr^{-1}}\right)\right] < -10$.

The possibility of magnetized WDs in AM CVn binaries has been explored previously  \citep{Roelofs2009, Farmer2010}. In the presence of a magnetized WD whose spin period is synchronized with the orbital period, disk winds could lead to magnetic braking, which could significantly accelerate the evolution of AM CVns -- especially at long periods, where gravitational wave angular momentum losses are weak. This might naturally explain the higher-than-predicted temperatures observed for most long-period AM CVn accretors \citep[e.g.][]{Ramsay2018, vanRoestel2022, Macrie2024}. We also note that all three of the systems we observed have donors that are reported to be inflated relative to predictions of models with gravitational wave-dominated evolution. This may not be a coincidence: if these systems have magnetized accretors, then magnetic braking will accelerate their evolution, allowing the the donors to retain higher entropy at the observed periods than expected without magnetic braking.  

Recently, \citet{Maccarone2024} showed that Gaia14aae and SDSS J080449.49+161624.8 (an AM CVn with $P_{\rm orb} = 44.5$ min) exhibit periodic X-ray variability that is modulated on the orbital period. They interpreted this as arising from magnetically collimated accretion and indicative of a magnetized accretor. This scenario requires the accretor's spin period to be synchronized with the orbital period, either through tides or through interaction of the accretor and donor magnetic fields \citep[e.g.][]{King1990}. Both tidal locking in WDs and the nature of interaction between the two components' magnetic fields -- if the donor has a magnetic field at all -- are poorly understood theoretically, but a synchronized and magnetized accretor is perhaps the simplest way to explain the observed X-ray periodicity in Gaia14aae. A magnetized accretor could similarly explain the hard X-ray spectrum and X-ray variability observed in SRGeJ0453 by \citetalias{Rodriguez2023}. 

The evidence we find for truncation of the inner disks provides additional evidence that moderate magnetic fields may be common in AM CVn accretors. The magnetic fields implied by Equation~\ref{eq:Beq} are typical of those found in isolated magnetic WDs \citep{Ferrario2015, Bagnulo2021} and are at least an order of magnitude weaker than those found in typical intermediate polars \citep[i.e., CVs with magnetically truncated disks;][]{1994patterson}. This reflects the fact that IPs have typical mass transfer rates a factor of $\sim 100$ lower than long-period AM CVns. That is, 100\,kG fields that can truncate the disks of long-period AM CVns would be undetectable in ordinary CVs. 

While magnetic disk truncation is a plausible explanation for the eclipse line profiles and lack of high-velocity emission we observe, our data alone do not definitively establish that the accreting WDs are magnetic. Studies of CVs and AM CVns in the optical have identified many other systems with emission line wings that are narrower than the Keplerian velocity at the surface of the WD accretor \citep[e.g.][]{Kaitchuck1994, Rau2010, Carter2013, Kupfer2016}. Most of these systems have not been reported to be magnetic, and some but not all display He II lines that are broader than their He I lines. The lack of high-velocity emission from the inner disk in many of these systems may thus be a result of radiative transfer effects that suppress emission in the inner disk rather than a genuine lack of disk material at small radii.  We conclude that the emission lines in Gaia14aae, SRGeJ0453, and ZTFJ1637 are consistent with magnetic disk truncation, but further data are needed to establish this as the most probable explanation. Searches for periodic X-ray variability in SRGeJ0453 and ZTFJ1637 are one promising avenue to further test the hypothesis of a magnetized accretor. 

\subsubsection{Lack of cyclotron features}
None of our targets show strong cyclotron features (``humps'') in their spectra.  Cyclotron humps are frequently observed in polars, which typically have surface magnetic fields exceeding 10 MG \citep[e.g.][]{1990cropper, 1994patterson}. The observed wavelengths of cyclotron humps constrains the magnetic field strength: setting the magnetic force equal to the centripetal force leads to the expression \citep[e.g.][]{Visvanathan1979}:
\begin{equation}
    \lambda = \frac{1.071\mu\textrm{m}}{n}\left(\frac{100\,\textrm{MG}}{B_{\rm WD}}\right)\sin\theta~, 
    \label{eq:cyc}
\end{equation}
where $n$ is the cyclotron harmonic number and $\theta$ is the angle between the magnetic pole and our line of sight. Cyclotron humps in polars are observed primarily in the optical; the strongest cyclotron humps are seen for $n=1-7$. At higher $n$, they become difficult to discern as individual features and blend into the continuum \citep[e.g.][]{2008campbell}. 

\begin{figure}
    \centering
    \includegraphics[width=0.5\textwidth]{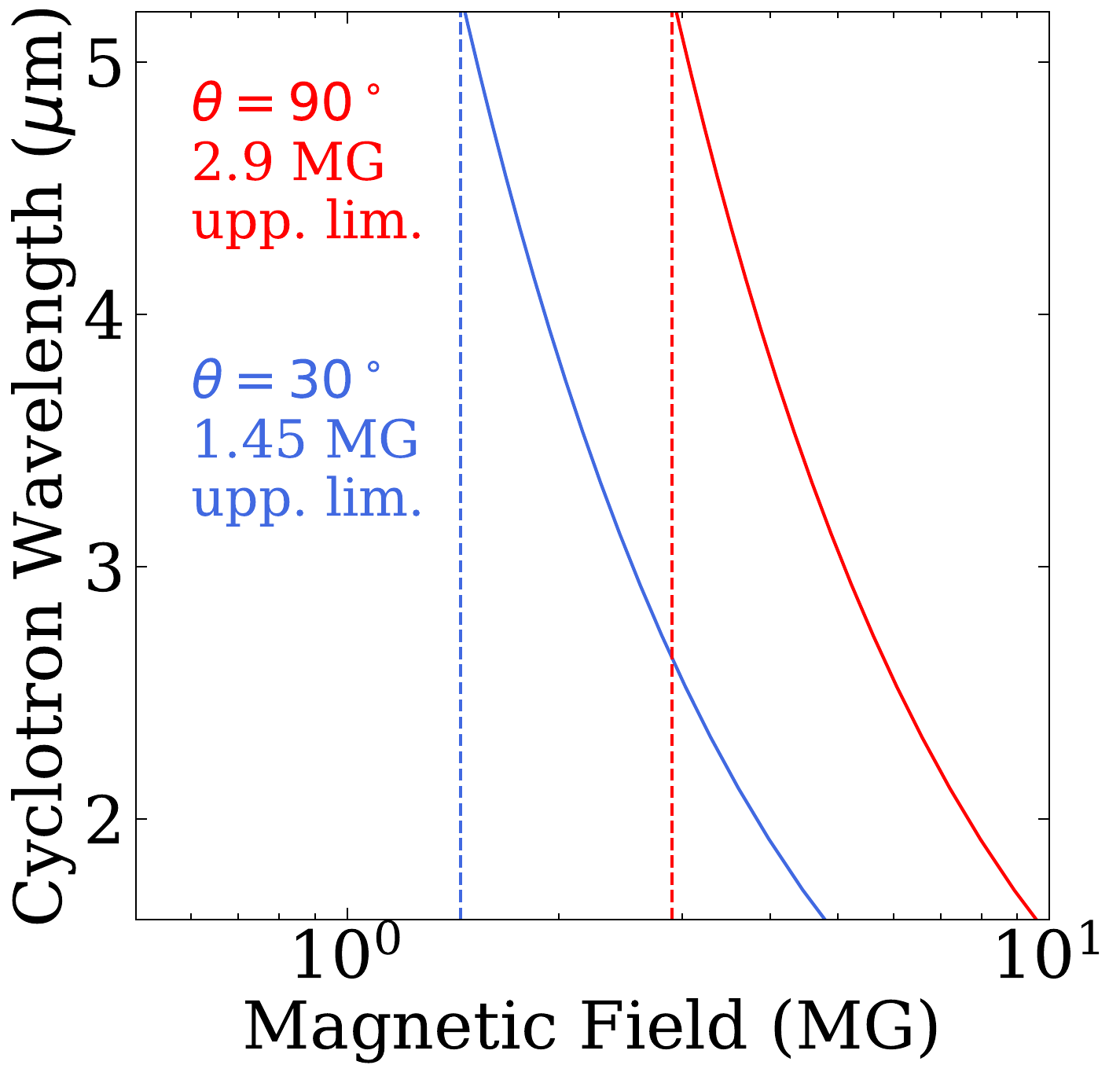}
    \caption{Upper limits on the WD magnetic field strength can be placed using Equation \ref{eq:cyc} and $n=7$. The lack of strong cyclotron features at the reddest wavelengths ($\approx5.2\mu$m) places strong upper limits on the magnetic field of the WD ($\approx2.9$ MG), assuming a magnetic pole inclination angle of 90$^\circ$. Choosing a smaller angle doesn't strongly affect this upper limit; unless magnetic pole angles are all very close to zero in all three systems, it is likely that these AM CVns have $\lesssim$ MG magnetic fields.}
    \label{fig:cyclotron}
\end{figure}

The lack of cyclotron humps in our IR spectra allows us to place upper limits on the WDs' magnetic fields that are somewhat more stringent than what could be set with optical data. Figure~\ref{fig:cyclotron} shows the predicted wavelength of the $n=7$ cyclotron hump for two choices of $\theta$. For $B_{\rm WD} \gtrsim 3\,{\rm MG}$, harmonics with $n\geq 7$ would produce cyclotron humps in the wavelength range probed by our data. For low inclination angles, the limits are somewhat lower. These upper limits are fully consistent with the lower limits inferred above from disk truncation: for plausible mass transfer rates of $\dot{M}\sim 10^{-11}\,M_{\odot}\,{\rm yr^{-1}}$, a magnetic field strong enough to produce observable cyclotron humps at $\lambda < 5\,{\rm \mu m}$ would dominate over ram pressure beyond the circularization radius, preventing the formation of a disk.

\subsection{Continuum flux from the donors}
\label{sec:thermal_donors}
We now consider the predicted emission from the donors, which depends sensitively on the uncertain donor temperatures. 

\subsubsection{Expected donor temperatures}
\label{sec:Teff_donor}

We assume that the donors are heated primarily via irradiation by the WD and accretion disk, which dominates over their internal heat \citep[e.g.][]{Deloye2007}. A minimum donor temperature can be calculated under the assumption that the donor reaches an equilibrium temperature where it uniformly radiates the energy intersected by its geometric cross section:

\begin{equation}
    \label{eq:irr}
    T_{{\rm uni}}=\left(1-A_{B}\right)^{1/4}\left(\frac{1}{4}\right)^{1/4}\sqrt{\frac{R_{{\rm wd}}}{a}}T_{{\rm eff,\,wd}}
\end{equation}
Here $A_B$ is the Bond albedo. Equation~\ref{eq:irr} assumes that the incident flux is uniformly re-radiated over the surface of the donor. It also neglects additional irradiation contributions from the accretion disk, which would increase $T_{{\rm uni}}$. A more general expression is $T_{{\rm uni}}=\left[\left(1-A_{B}\right)L_{{\rm tot}}/\left(16\pi\sigma a^{2}\right)\right]^{1/4}$, where $L$ is the total luminosity of the WD and accretion flow, and $\sigma$ is the Stefan-Boltzmann constant.

To parameterize the efficiency of heat redistribution in the donor, we introduce a heat redistribution efficiency factor, $\varepsilon$ \citep[e.g.][]{Cowan2011}, which varies from 0 to 1. If heat redistribution is inefficient ($\varepsilon = 0$), the apparent day-side effective temperature of the donor is $T_{{\rm day},\,\varepsilon=0}=\left(\frac{8}{3}\right)^{1/4}T_{{\rm uni}}$, which is higher by a factor of $\approx 1.28$ than the expected temperature in the case of uniform heat redistribution. In this limit, the night-side of the donor is much colder, with $T_{\rm night,\,\varepsilon = 0}\approx 0$. 

Following \citet{Cowan2011}, we interpolate the predicted day-side apparent temperature at intermediate $\varepsilon$ as 
\begin{equation}
    \label{eq:Tirr_eps}
    T_{{\rm day}}=T_{{\rm uni}}\left(\frac{8}{3}-\frac{5}{3}\varepsilon\right)^{1/4}.
\end{equation}
The efficiency factor $\varepsilon$ is defined such that Equation~\ref{eq:Tirr_eps} holds. The corresponding night-side temperature is 

\begin{equation}
    \label{eq:Tnight}
    T_{{\rm night}}=T_{{\rm uni}}\varepsilon^{1/4}.
\end{equation}

Although $A_B$ and $\varepsilon$ are uncertain for AM CVn donors, they have been constrained for many brown dwarfs and hot Jupiters \citep[e.g.][]{Schwartz2015, Lew2022, Zhou2022, Blazek2022, Amaro2023, Lothringer2024, Amaro2024, French2024, Casewell2024, Amaro2025}. Weakly irradiated brown dwarfs and hot Jupiters are found to have efficient energy circulation, with only modest day--night asymmetries in their phase curves. More strongly irradiated brown dwarfs display stronger day--night brightness contrasts, suggesting less efficient energy redistribution. \citet{Amaro2025} find that the transition between these two regimes occurs at an irradiation flux of $F_{\rm irr}\approx 10^9\,{\rm erg\,s^{-1}\,cm^{-2}}$.

All of our targets have $F_{\rm irr}$ near or below $10^9\,{\rm erg\,s^{-1}\,cm^{-2}}$ (Table~\ref{tab:sample}), so we assume they are in the weakly irradiated regime.  Most observed weakly irradiated brown dwarfs and hot Jupiters also have low inferred albedos, consistent with predictions of atmosphere models \citep[e.g.][]{Marley1999, Sudarsky2000, Morley2015}. In Table~\ref{tab:sample}, we assume $\varepsilon=0.8$ and $A_B=0$. This results in predicted $T_{\rm uni}$ ranging from 1000 to 1700\,K and modest $\sim 10\%$ temperature variations between the predicted day- and night-side temperatures.

\begin{figure}
    \centering
    \includegraphics[width=\columnwidth]{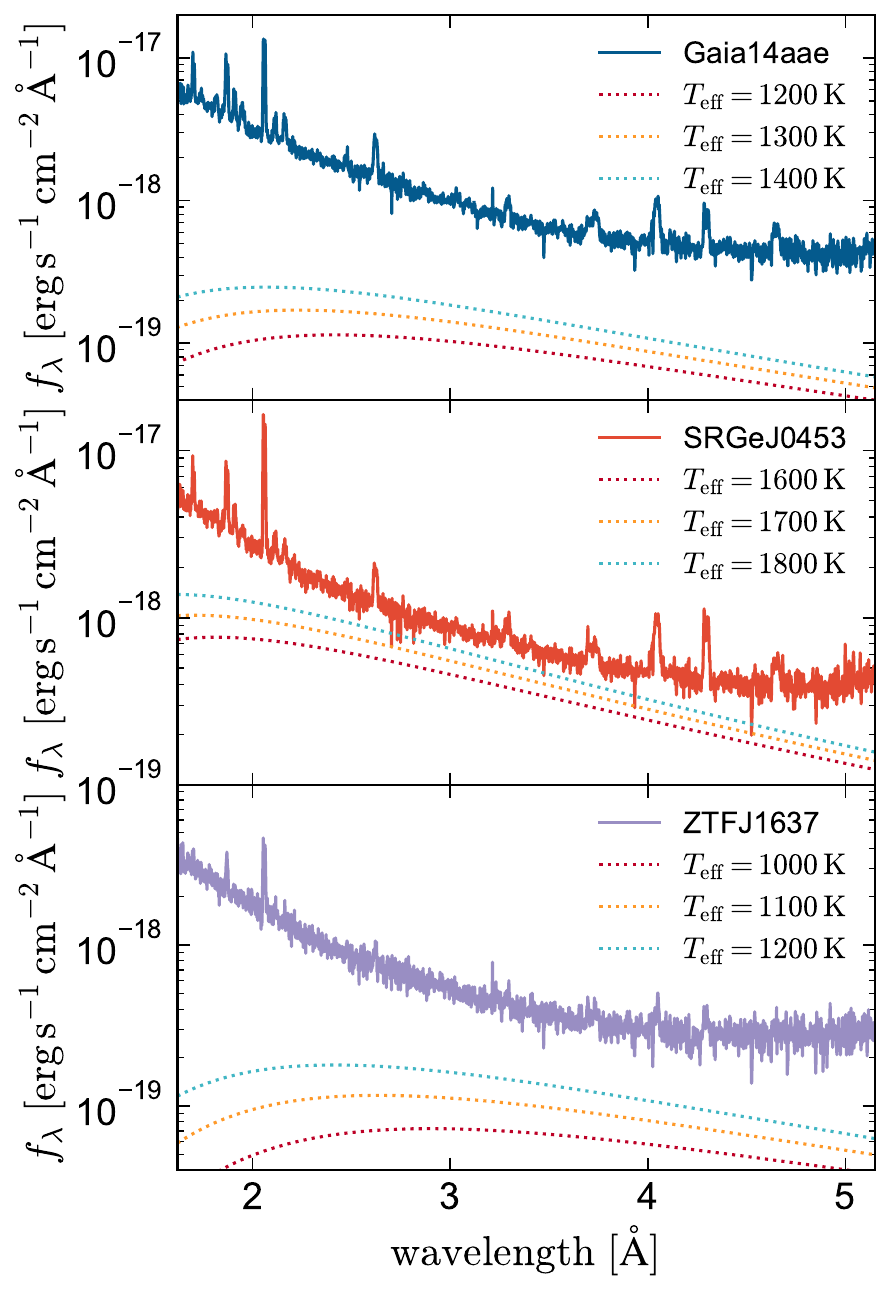}
    \caption{Comparison of the phase-averaged spectra of the three observed AM CVns to blackbody spectra with $T_{\rm eff}$ bracketing the donor temperatures expected from irradiation equilibrium.  }  
    \label{fig:phoenix}
\end{figure}

\subsubsection{Comparison to blackbody models of the donors}
Figure~\ref{fig:phoenix} compares the observed phase-averaged spectra of the three systems to models for the donor SEDs with a range of $T_{\rm eff}$. We model the donor SEDs as blackbodies with radii set to the values in Table~\ref{tab:sample} and effective temperatures corresponding approximately to $T_{\rm night}$, $T_{\rm uni}$, and $T_{\rm day}$ reported in Table~\ref{tab:sample2}.

Blackbody models predict that the donor should contribute $\gtrsim 50\%$ of the observed flux in SRGeJ0453 at $2.5-3.5\,{\rm \mu m}$, with its relative flux contribution falling at shorter and longer wavelengths. For the other systems, the blackbody models with $T_{\rm eff} \approx T_{\rm uni}$ predict the donors to contribute $\lesssim 20\%$ of the continuum flux. If the donors have sharp absorption features in their spectra, they should be detectable and strongly RV-variable at these flux ratios, like the Na\,I lines. Setting quantitative upper limits on the donor temperatures will require spectral models for helium-rich low-temperature donors, which do not currently exist.

The lack of detected secondary eclipses (Figure~\ref{fig:lightcurves}) disfavors scenarios in which the donor both contributes more than a few percent of the observed flux and is efficiently occulted by the disk. However, since the outer disk may be geometrically or optically thin in the near-IR, the lack of a detected secondary eclipse does not translate to a simple constraint on the donor temperature.

\citetalias{vanRoestel2022} reported that the IR excess of ZTFJ1637 in the {\it WISE} $W_1$ and $W_2$ bands can be fully explained by thermal emission from the donor, which they modeled as a blackbody. However, we find that the blackbody with $T_{\rm eff} = T_{\rm uni}$ (Equation~\ref{eq:irr}) underestimates the flux at 3.4 and 4.6 by a factor of several. We conclude that most of the infrared excess at $2-5\mu$m comes from the disk, not the donor.

\section{Conclusions}
\label{sec:conclusions}
We have presented high-cadence, phase-resolved spectroscopy of three eclipsing AM CVns binaries at $\lambda = 1.6-5.2\,{\rm \mu m}$. Our targets have orbital periods of 50, 55, and 61 minutes, placing them in the regime where outbursts are rare, the donor is cold, and optical photometry is dominated by the accreting WD. All our targets have IR excesses that were proposed to be the result of thermal emission from the donors. Our data reveals that the excess is dominated by the systems' cold accretion disks, not the donors, and we do not detect any unambiguous absorption lines in the data. We do, however, detect the donors of two systems in emission. Our main results are as follows:

\begin{enumerate}
    \item {\it Disk-dominated spectra}: All three systems have spectra dominated by emission lines, most of which are double-peaked (Figures~\ref{fig:phase_avg_with_wd}, \ref{fig:trailed_235m}, and \ref{fig:trailed_395m}). The accreting WDs contribute about half the total light at 1.6\,$\mu$m, but only $\sim 10\%$ at 5\,$\mu$m. Most lines are not detectably RV variable in their centers, but several exhibit line shape variability. 

    \item {\it Eclipse of the WD and disk}: Our observations cover an eclipse of the WD and disk in each system, including two total eclipses and one grazing eclipse. Subtracting the in-eclipse spectrum from the out-of-eclipse spectrum reveals the spectrum of the WD accretor (Figure~\ref{fig:eclipse}). The disk eclipses last $\sim 5$ times longer than the WD eclipses and are most visible at long wavelengths (Figure~\ref{fig:lightcurves}). For two of the three systems, the donor's trajectory across the disk is clearly visible in position-velocity space (Figure~\ref{fig:trailed_HeI}); the shape of this  trajectory constrains the geometry of the disk and donor (Figure~\ref{fig:sim_spectrogram}). 

    \item {\it RV-variable components in the disk}: The strong emission lines of all three systems contain components that are RV-variable and likely trace the ``bright spot'', where the accretion stream intersects the disk (Figure~\ref{fig:trailed_HeI}). In ZTFJ1637, the bright spot's RV curve is measured sufficiently well to provide a constraint on the mass ratio and disk radius that is independent  of constraints from the light curve (Figure~\ref{fig:bright_spot}). In the other two systems, there is evidence of two bright spots, perhaps due to a second impact spot of the stream on the other side of the disk (Figure~\ref{fig:doppler_2.058}).

    \item {\it Sodium emission tracing the donor}: In Gaia14aae and SRGeJ0453, two sets of sodium emission lines are strongly RV variable and follow the predicted RV curve of the donor (Figure~\ref{fig:trailed_Na}). The lines are bright when the accretor is in front of the donor, and absent when the donor is in front of the accretor. We interpret this as emission from the irradiated side of the donor. 

    \item {\it Possible evidence for disk truncation}: We develop a simple model to interpret trailed spectrograms of disk emission lines during eclipse (Figure~\ref{fig:sim_spectrogram}). We find that both the morphology of the line eclipse and the low velocities of the line wings suggest that the disk does not extend to the WD surface, but is truncated between $0.05$ and $0.1\,R_\odot$. We conjecture that this is a consequence of a magnetic WD accretor, with an implied surface field of $30-100$\,{\rm kG}.

    \item {\it Limits on contributions from the donor}: Besides the Na\,I lines on the donors' irradiated faces, we do not detect any spectral features from the donors, and we do not detect a secondary eclipse when the donor is occulted by the accretor and disk. If the donors are in thermal equilibrium and heated only by irradiation from the disk and WD, they are predicted to contribute $\approx 10-50\%$ of the flux in our observations, depending on their albedo and atmospheric circulation efficiencies (Figure~\ref{fig:phoenix}). Since the SEDs of the donors likely differ significantly from those of hydrogen-rich objects, obtaining more precise limits on the donor flux contributions will require bespoke atmosphere models for cold, degenerate, helium-rich objects. 
    
\end{enumerate}

Our results highlight several open questions that can be addressed with future {\it JWST} observations. A broader sample of long-period AM CVns is needed to determine whether donor-line detections and truncated disks are common or reflect system-to-system variations. Higher SNR, multi-orbit spectroscopy -- especially of systems with well-constrained masses--would clarify the origin of the different He I line morphologies and enable more robust constraints on the inner disk radius. The tentative evidence we find for weak magnetic fields in the accreting WDs motivates further investigation of disk–field interactions, spin-orbit coupling, and  magnetic braking in long-period AM CVns. Finally, spectroscopic detection of AM CVn donors -- thus far only in emission -- opens the possibility of direct measurements of donor temperatures, albedos, and atmospheric compositions, providing new ways to discriminate among formation channels. Fully exploiting this data will require the development of bespoke atmosphere models for cool, helium-dominated, irradiated objects. We defer such modeling to future work. 

\section*{acknowledgments}
We are grateful to the anonymous referee for a constructive report. We also thank Tom Maccarone, Jan van Roestel, Peter Hauschildt, and Sunny Wong for useful discussions.

This work is based on observations made with the NASA/ESA/CSA James Webb Space Telescope. The data were obtained from the Mikulski Archive for Space Telescopes at the Space Telescope Science Institute, which is operated by the Association of Universities for Research in Astronomy, Inc., under NASA contract NAS 5-03127 for JWST. These observations are associated with program \#04979.

This research was supported by JWST-GO-04979.001-A and by NSF grants AST-2307232 and AST-2508988. This research was supported in part by grant NSF PHY-2309135 to the Kavli Institute for Theoretical Physics (KITP).

\newpage

\newpage

\bibliographystyle{mnras}



\end{document}